%% file: paper.tex
\pdfoutput=1
\documentclass[newfonts=false,format=sigconf,10pt,letterpaper]{acmart}

\usepackage[english]{babel}

\usepackage{xcolor}
\usepackage{wrapfig}

\newcommand{\ankitnew}{\textcolor{black}}
\newcommand{\gregnew}{\textcolor{black}}

\usepackage{numprint}
\usepackage{blindtext, graphicx}
\usepackage{algorithm}
\usepackage{algpseudocode}

\newcolumntype{L}[1]{>{\raggedright\let\newline\\\arraybackslash\hspace{0pt}}m{#1}}
\newcolumntype{C}[1]{>{\centering\let\newline\\\arraybackslash\hspace{0pt}}m{#1}}
\newcolumntype{R}[1]{>{\raggedleft\let\newline\\\arraybackslash\hspace{0pt}}m{#1}}

\usepackage{xspace}

\newcommand{\ie}{{\it i.e.,}\xspace}
\newcommand{\eg}{{\it e.g.,}\xspace}

\usepackage{subfigure}
\usepackage{enumitem}

\newcommand{\parab}[1]{\vspace{0.025in}\noindent\textbf{#1}}

\usepackage{siunitx}
\newcommand{\figcap}[1]{\caption{\textit{#1}}}

\renewcommand\footnotetextcopyrightpermission[1]{} 
\setcopyright{none}

\acmDOI{}

\acmISBN{}

\acmPrice{}

\begin{document}
\title{cISP: A Speed-of-Light Internet Service Provider}



\author{\fontsize{10pt}{12pt}\selectfont Debopam Bhattacherjee$^1$, Sangeetha Abdu Jyothi$^2$, Ilker Nadi Bozkurt$^3$, Muhammad Tirmazi$^4$, Waqar Aqeel$^3$, Anthony Aguirre$^5$, Balakrishnan Chandrasekaran$^6$, P. Brighten Godfrey$^2$,  Gregory Laughlin$^7$, Bruce Maggs$^{3,8}$, Ankit Singla$^1$\\\vspace{8pt}
$^1$ETH Z\"{u}rich, $^2$UIUC, $^3$Duke University, $^4$LUMS, $^5$UCSC, $^6$MPI Saarbr\"{u}cken, $^7$Yale University, $^8$Akamai Technologies\vspace{0.15in}}


\begin{abstract}
Low latency is a requirement for a variety of interactive network applications. The Internet, however, is not optimized for latency. 
We thus explore the design of cost-effective wide-area networks that move data over paths very close to great-circle paths, at speeds very close to the speed of light in vacuum. Our \textbf{cISP} design augments the Internet's fiber with free-space wireless connectivity.
  cISP addresses the fundamental challenge of simultaneously providing low latency and scalable bandwidth, while accounting for numerous practical factors ranging from transmission tower availability to packet queuing. 
We show that instantiations of cISP across the contiguous United States and Europe would achieve mean latencies within $5\%$ of that achievable using great-circle paths at the speed of light, over medium and long distances. Further, we estimate that the economic value from such networks would substantially exceed their expense.
\end{abstract}

\maketitle

\input{intro}

\input{background}
\input{microwaveDesign}
\input{queuing}

\input{practical}
\input{casestudies}
\input{costbenefit}
\input{related}
\input{conclusion}

\bibliographystyle{ACM-Reference-Format}
\bibliography{cISP}

\end{document}

%% file: intro.tex
\section{Introduction}
\label{sec:intro}

User experience in many interactive network applications depends crucially on achieving low latency.
Even seemingly small increases in latency can negatively impact user experience, and, subsequently, revenue for the service providers: Google, for example, quantified the impact of an additional $400$~ms of latency in search results as $0.7\%$ fewer searches per user~\cite{googSearch}. Further, wide-area latency is often the bottleneck, as Facebook's analysis of over a million requests found~\cite{fbLatency}.  Indeed, content delivery networks present latency reduction and its associated increase in conversion rates as one of the key value propositions of their services, citing, \eg a $1\%$ loss in sales per $100$~ms of latency for Amazon~\cite{akamai10for10}.
In spite of the significant impact of latency on performance and user experience, the Internet is not designed to treat low latency as a primary objective.
This is the problem we address: reducing latencies over the Internet to the lowest possible.

The best achievable latency between two points along the surface of the Earth is determined by their geodesic distance divided by the speed of light, $c$. Latencies over the Internet, however, are usually much larger than this minimal ``$c$-latency'': recent measurement work found that fetching even small amounts of data over the Internet typically takes $37\times$ longer than the $c$-latency, and often, more than $100\times$ longer~\cite{soslow-pam17}. This delay comes from the many round-trips between the communicating endpoints, due to inefficiencies in the transport and application layer protocols, and from each round-trip itself taking $3$-$4\times$ longer than the $c$-latency~\cite{soslow-pam17}. Given the approximately multiplicative role of network round-trip times (RTTs) (when bandwidth is not the main bottleneck), eliminating inflation in Internet RTTs can potentially translate to up to $3$-$4\times$ speedup, even without any protocol changes. Further, as protocol stack improvements get closer to their ideal efficiency of one RTT for small amounts of data, the RTT becomes the singular network bottleneck. Similarly, for well-designed applications dependent on persistent connectivity between two fixed locations, such as gaming, \emph{nothing} other than resolving this $3$-$4\times$ ``infrastructural inefficiency'' can improve latency substantially.

Thus, beyond the networking research community's focus on protocol efficiency, reducing the Internet infrastructure's latency inflation is the next frontier in research on latency. While academic research has typically treated infrastructural latency inflation as an unresolvable given, we argue that this is a high-value opportunity, and is much more tractable than may be evident at first. It is even plausible that infrastructural improvements will be easier and faster to deploy than our ongoing decades-long efforts towards new protocols.



What are the root causes of the Internet's infrastructural inefficiency, and how do we ameliorate them? Large latencies are partly explained by poor use of existing fiber infrastructure: two communicating sites often use a longer, indirect route because their service providers do not peer over the shortest fiber connectivity between their locations. We find, nevertheless, that even latency-optimal use of \emph{all} known fiber conduits, computed via shortest paths in the recent InterTubes dataset~\cite{intertubes}, would leave us $1.93\times$ away from $c$-latency. This gap stems from the speed of light in fiber being $\sim$$\frac{2}{3}c$, and the unavoidable circuitousness of fiber routes due to topographic and economic constraints of buried conduits.

We thus explore the design of \textbf{cISP}, an Internet Service Provider that provides nearly speed-of-light latency by exploiting wireless electromagnetic transmissions, which can be realized with point-to-point microwave antennae mounted on towers.  This approach holds promise for overcoming both the aforementioned shortcomings fundamental to today's fiber-based networks: the transmission speed in air is essentially equal to $c$, and the richness of existing tower infrastructure makes more direct paths possible. Nevertheless, it also presents several new challenges, including:

\begin{itemize}
\setlength\itemsep{1pt}
    \item overcoming numerous practical constraints, including tower availability, line-of-sight requirements, and the impact of weather on performance;
    \item coping with the limited wireless bandwidth;
    \item solving a large-scale cost-optimal network design problem, which is NP-hard; and
    \item addressing switching and queuing delays, which are more prominent with the smaller propagation delays. 
\end{itemize}


To meet these challenges, we propose a hybrid design that augments the Internet's fiber connectivity with nearly straight-line wireless links. These low-latency links are used judiciously where they provide the maximum latency benefit, and only for the small proportion of Internet traffic that is latency-sensitive. We design a simple heuristic that achieves near-optimal results for the network design problem. Our approach is flexible and enables network design for a variety of deployment scenarios; in particular, we show that cISP's design for interconnecting large population centers in the contiguous U.S.\ and Europe can achieve mean latencies as low as $1.05\times$ $c$-latency at a cost of under $\$1$ per gigabyte (GB). We show through simulation that such networks can be operated at high utilization without excessive queuing.

To address the practical concerns, we use fine-grained geographic data and the relevant physical constraints to determine where the needed wireless connectivity would be feasible to deploy, and assess our design under a variety of scenarios with respect to budget, tower height and availability, antenna range, and traffic matrices. We also use a year's worth of meteorological data to assess the network's performance during weather disturbances, showing that most of cISP's latency benefits remain intact throughout the year. 

Our weather simulation and an animation showing how the hybrid network evolves from mostly-fiber to mostly-wireless with increasing budget are available online; see \cite{anonWeatherVizLink} and \cite{anonNetworkAnim}. 

Lastly, we explore the application-level benefits for Web browsing and gaming, and present estimates showing that the utility of cISP vastly exceeds its cost.

%% file: background.tex
\section{Technology background}
\label{sec:background}


At the highest level, our approach involves using free-space communication between transmitters mounted at a suitable height, \eg using dedicated towers or existing buildings, and separated from each other by at most a certain limiting distance. Network links longer than this range require a series of such transmitters. Typically, even after accounting for terrain, such network links can be built close to the shortest path on the Earth's surface between the two end points. Further, the speed of light in air is essentially the same as that in vacuum, $c$. These properties make this approach attractive for the design of (nearly) $c$-latency networks.

\parab{Technology choices.} Several physical layer technologies are amenable for use in our design, including free-space optics (FSO), microwave (MW), and millimeter wave (MMW). At present, we believe MW provides the best combination of range, resilience, throughput, and cost. Future advances in any of these technologies, however, can be easily rolled into our design, and can only improve our cost-benefit analysis.

While hollow fiber~\cite{hollowFiber} could, in the future, also provide $c$-latency, it would still suffer from the circuitousness of today's fiber conduits. Low-Earth-orbit satellites may also help, but their connectivity fundamentally varies over time, necessitating extremely high density to provide latencies similar to those achievable with a terrestrial MW network. 

\parab{Switching latency.} While long-haul MW networks have existed since the 1940s~\cite{attLongLines}, their recent use in high-frequency trading has driven innovation in  radios so that each MW retransmission only takes a few $\mu s$. Thus, even wide-area links with many retransmissions incur negligible switching latency. As an example, the HFT industry operates a MW relay between New Jersey and Chicago comprising $\sim20$ line-of-sight links that operates within $1\%$ of $c$-latency end-to-end at the \emph{application} layer~\cite{McKayLatency2017}.

\parab{Packet loss.} Packet loss occurs for several reasons, including, notably, weather disruption, and intermittent multi-path fading, especially over bodies of water.

To examine loss, we obtained network performance data for an FCC-licensed, operational Chicago-to-New-Jersey MW relay link from an ultra-low latency wireless provider. The data comprise $2$,$743$ distinct one-minute intervals between 10/22/2012 through 11/01/2012, spanning the intervals 9:30AM -- 4:00PM EDT when both the futures and equity markets were open and trading in Aurora, IL, and Carteret, NJ, respectively. This period provides an extreme test: Hurricane Sandy caused broad, serious disruption in the NJ area for $4$ days within this period. These data thus show a high average packet loss rate of $16.1\%$. Even for this period, the median loss rate is much smaller at $1.4\%$. In Section~\ref{section:weather} we present a broader analysis of the impact of diverting traffic to alternate (fiber or MW) routes during inclement weather using a year's worth of weather data.

Note also that the IL-NJ link was designed to absolutely minimize latency for the HFT industry. Hence forward error correction spanning multiple packets was used minimally or not at all. A less aggressive design using such techniques, together with shorter tower-to-tower distances is likely to further reduce loss rates.


\parab{Spectrum and licensing.} We propose the use of MW communication in the $6$-$18$~GHz frequency range. These frequencies are not very crowded, and licensing is generally not very competitive, except at $6$~GHz in cities, and along certain routes, like the above mentioned HFT corridor. The licenses are given on a first-come, first-served basis, recorded in a public database, and they protect against the deployment of other links that would interfere with licensed links.

\parab{Line-of-sight and range.} Successive MW towers need line-of-sight visibility, accounting for the Earth's curvature, terrain, trees, buildings, and other obstructions, and atmospheric refraction. 
Attenuation also limits range. A maximum range of around $100$~km is practicable, but we show results with maximum allowed range varying between $60$-$100$~km (\S\ref{subsec:heightAvailability}).




\parab{Bandwidth.} Between any two towers, using very efficient encoding (256 QAM or higher), wide frequency channels, and radio multiplexing, a data rate of about $1$~Gbps is achievable~\cite{hansryd2011microwave}.
This bandwidth is vastly smaller than for fiber, and necessitates a hybrid design using fiber and MW.


\parab{Geographic coverage.} 
Connecting individual homes directly to such a MW network would be cost-prohibitive. To maximize cost-efficiency, we focus on long-haul connectivity, with the last mile being traditional fiber.
At short distances, fiber's circuitousness and refraction are small overheads.

\parab{Cost model.} We rely on cost estimates in recent work~\cite{cost} and based on our conversations with industry participants involved in equipment manufacturing and link provisioning. The cost of installing a bidirectional MW link, on existing towers, is approximately $\$75$K ($\$150$K) for $500$~Mbps ($1$~Gbps) bandwidth. The average cost for building a new tower is $\$100$K, with wide variation by terrain and across cities and rural areas. Any additional towers needed to augment bandwidth for particular links incur this ``new tower'' cost. 

The operational costs comprise several elements, including management and personnel, but the dominant operational expense, by far, is tower rent: $\$25-50$K per year per tower. We estimate cost per GB by amortizing the sum of building costs and operational costs over $5$ years.





%% file: microwaveDesign.tex
\section{\lowercase{c}ISP Design}
\label{sec:microwaveDesign}

At an abstract level, given the tower and fiber infrastructure, a set of sites (\eg cities, data centers) to interconnect, and a traffic model between them, we want to select a set of tower-level connections that minimize network-wide latency while adhering to a budget and the constraints outlined in \S\ref{sec:background}. Our approach comprises the following three broad steps.

\begin{enumerate}\setlength\itemsep{2pt}
    \item Identifying a set of links that are likely to be useful by determining, for each pair of sites ($s$, $d$), the best feasible tower-level connectivity, if $s$ and $d$ were to be directly connected by a series of towers.
    \item Building all $O(n^2)$ direct links, connecting each site to every other, would be prohibitively expensive. Thus, a subset of site-to-site links, together with existing fiber conduits, form our network. Choosing the appropriate subset is the key algorithmic problem. 
    \item Provisioning capacity beyond $1$~Gbps along any link involves building additional tower-level links, \eg by identifying and using links that are also nearly shortest paths, but were omitted in step 1 above.
\end{enumerate}

\subsection{Step 1: Feasible hops}

We first use line-of-sight and range constraints to decide which tower pairs can be connected. Achievable tower-to-tower hop length is limited primarily by the Earth's curvature. MW hops must clear this curvature and any obstructions in an ellipsoidal region between the sender and the receiver antennae called the Fresnel zone with width $h_{\rm Fres}$. The Earth's curvature can be treated as a ``bulge'' of height $h_{\rm Earth}$ that a straight-line path must
clear.  At the midpoint of a hop of length $D$, using a MW frequency $f$, we have:

\vspace{-.15in}
$$
h_{\rm Fres} \simeq 8.7\,m\left(D\over 1\,{\rm km}\right)^{1/2}\left({f\over 1\,{\rm GHz}}\right)^{-1/2}
$$
$$
h_{\rm Earth} \simeq {1~m\over {50~K}}\left({D\over 1\,{\rm km}}\right)^2
$$
where $K$ accounts for atmospheric refraction~\cite{Manning}.  
Towers should clear the sum of these heights and any other obstructions. In favorable weather, and with adequately large dish antennae, ranges of up to $D\approx100$~km are achievable at high availability, provided such line-of-sight clearance~\cite{shkilko}. As a specific example, the FCC licensing database~\cite{FCCdatabase} 
indicates that McKay Brothers, LLC (a provider for the financial industry) operated a $D=96\,{\rm km}$ hop from Chicago, IL (lat. 41.88$^{\circ}$, lon. -87.62$^{\circ}$) to Galien, MI (lat. 41.81$^{\circ}$, lon. -86.47$^{\circ}$) as part of a
$1183\,{\rm km}$ MW relay. This shows that multipath interference issues (associated in this case with a traversal over Lake Michigan) are not an impediment to hop viability.


We assess hop feasibility between each pair of towers by using terrain data made available by NASA~\cite{terrain_data}, 
which includes buildings and ground clutter, and effectively incorporates the height of the tree canopy.\footnote{\gregnew{This NASA data set combines data from the Shuttle Radar Topography Mission (SRTM)~\cite{terrain_data} and the National Elevation Database (NED)~\cite{ned_data}, and typically yields acceptably small error ($\sim$$2$~m) against reference, high-accuracy LIDAR measurements.}} We also require a fully clear Fresnel zone, and adopt $K=1.3$ and $f=11\,$ GHz in the above formulae. 
We have used our hop engineering routines to design line-of-sight networks, at least $4$ of which are now deployed, including ultra-low latency routes between data centers hosting financial market matching engines. Our methodology has routinely provided correct clearance assessments when the physical paths are flashed. It is relatively rare that the hop feasibility assessment is inaccurate; if a problem arises, it is most likely that the locations themselves are not available to rent. In \S\ref{subsec:heightAvailability}, we explore relaxations of the tower rental assumptions.

After identifying feasible tower-to-tower hops, for each pair of sites, we find the shortest path through a graph containing these hops, which we call a {\em link}. In line with observations from the tower data around major population centers, we assume each site itself hosts enough towers to use as the starting point for connectivity from that site to many others.



\subsection{Step 2: Topology design}
\label{sec:topo-design}

Picking a subset of these site-to-site links involves solving a typical network design problem. The Steiner-tree problem~\cite{steinerNP} can be easily reduced to this problem, thereby establishing hardness. However, standard approximation algorithms, like linear program relaxation and rounding, yield sub-optimal solutions, which although provably within constant factors of optimal, are insufficient in practice. We develop a simple heuristic, which, by exploiting features specific to our problem setting, obtains nearly optimal solutions.

\parab{Inputs:} Our network design algorithm requires:
\begin{itemize}
\setlength\itemsep{1pt}
    \item A set of sites to be interconnected, $v_1, v_2, \ldots, v_n$.
    \item A traffic matrix $H$ specifying the relative traffic volume $h_{ij} \in [0,1]$ between each pair $v_i$ and $v_j$.
    \item The geodesic distance $d_{ij}$ between each $v_i$ and $v_j$.
    \item The distance along the shortest, direct MW path between each pair, $m_{ij}$, as well as its cost, $c_{ij}$. This is part of the output of step 1.
    \item The optical fiber distance between each pair, $o_{ij}$, which we multiply by $1.5$ to account for fiber's higher latency.
    \item A total budget $B$ limiting the maximum number of bidirectional MW links that can be built.
\end{itemize}

\parab{Expected output:} The algorithm must decide which direct MW links to pick, \ie assign values to the corresponding binary decision variables, $x_{ij}$, such that the total cost of the picked links fits the budget, \ie $\sum_{ij} x_{ij}c_{ij} \leq B$. Our objective is to minimize, per unit traffic, the mean stretch, \ie the ratio of latency to $c$-latency, where $c$-latency is the speed-of-light travel time between the source and destination of the traffic. 

\parab{Problem formulation:} Expressing such problems in an optimization framework is non-trivial: we need to express our objective in terms of shortest paths in a graph that will itself be the \emph{result}. We use a formulation based on network flows.

Each pair of sites ($v_s$, $v_t$) exchanges $h_{st}$ units of flow. To represent flow routing, for each potential link $\ell$, we introduce a binary variable $f_{stij,m}$ which is $1$ \emph{iff} the $v_s$$\rightarrow$$v_t$ flow is carried over the microwave link $v_i$$\rightarrow$$v_j$, and a binary variable $f_{stij,o}$ which is $1$ \emph{iff} the same flow is carried over the optical link\footnote{A ``link'' between sites can use multiple physical layer hops, both for MW and fiber. The underlying multi-physical-hop distances are already captured by the inputs $o_{ij}$ and $m_{ij}$ so the optimization views it as a single link.} $v_i$$\rightarrow$$v_j$. The objective function is:
\begin{equation}
\label{eq_obj}
min \sum_{s,t} \frac{h_{st}}{d_{st}} \,\sum_{i,j}\left( o_{i,j} f_{stij,o} + m_{i,j} f_{stij,m}\right)
\end{equation}
The $h_{st}$ term achieves our goal of optimizing \emph{per unit traffic}. The $\frac{1}{d_{st}}$ term achieves our goal of optimizing the \emph{stretch}.

For brevity, we omit the constraints, which include: flow input and output at sources and sinks; flow conservation; total budget; and the requirement that only links that are built ($x_{ij}=1$) may carry flow. All variables are binary, so flows are ``unsplittable'' (carried along a single path) and the overall problem is an integer linear program (ILP).

Note that we have decomposed the problem so that link capacity is \emph{not} a constraint in this formulation: MW links will be built with sufficient capacity in step 3; fiber links are assumed to have plentiful bandwidth at negligible cost relative to MW costs. As a result, the objective function will guide the optimizer to direct each $v_i \to v_j$ flow along the shortest path of built links, which is the direct MW link $v_i \to v_j$ if it happens to be built, or otherwise, a path across some mix of one or more fiber and MW links.

\parab{Solution approach:} As we shall see, simply handing the ILP to a solver did not scale to beyond medium-sized networks.  By exploiting our problem structure, however, we develop a simple heuristic that yields near-optimal results at smaller scales (verified against the exact ILP solution) and can solve the problem at the larger scales of interest.

The first observation we make is that a large number of variables in our formulation will never take non-zero values, allowing us to eliminate them and any resulting null constraints. Roughly stated: if, for a particular ($v_s$,$v_t$) pair, a microwave path is of higher latency than a fiber path (which we can always use, at zero expense), then it will never carry $v_s$$\rightarrow$$v_t$ flow, though other flows may still traverse it. Similar observations apply to individual ``distant, off-path'' fiber and MW links. This simple observation substantially reduces the problem size. Note that standard network design problems do not typically have this structure available. This is entirely due to the hybrid design using fiber, which is assumed to be cheap, where available. We benefit, in this case, from having an ``oracle'' that tells us \emph{a priori} when certain flow assignments are ``obviously bad'' and will not be useful. Further, carefully defined, such constraints preserve optimality; this part of our solution is not an approximation.

Second, we use a fast greedy heuristic to prune out MW links that are unlikely to be chosen. The heuristic operates using a larger budget ($2\times$ in our implementation) than we are ultimately allowed.  In each iteration, we add to the solution the MW city-to-city link that decreases average stretch the most, continuing until the total cost reaches the inflated budget; the chosen links are candidates given to the ILP. Intuitively, the other links are uninteresting -- they are unlikely to be picked in the final optimization even when a substantially larger budget is available, and so are not presented as options to the ILP. This approach does not provide any guarantees, but we find that on small problem sizes, where the exact ILP can also be evaluated, it obtains the optimal solution.


\begin{figure}
\centering
\includegraphics[height=2in]{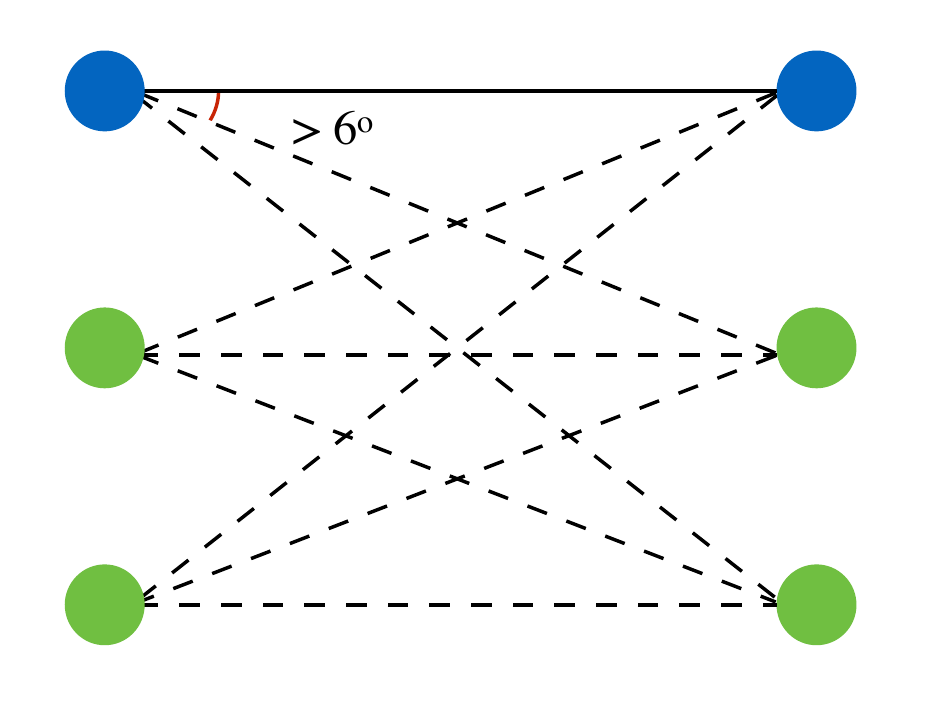}
\vspace{-0.1in}
\caption{ \em Bandwidth augmentation: $k^2$ hops with $O(k)$ new towers. }
\label{fig:bw-augment}
\end{figure}

\begin{figure*}[t]
  \centering
  \subfigure[]{\label{fig:ilp:runtime}
    \includegraphics[width=3in]{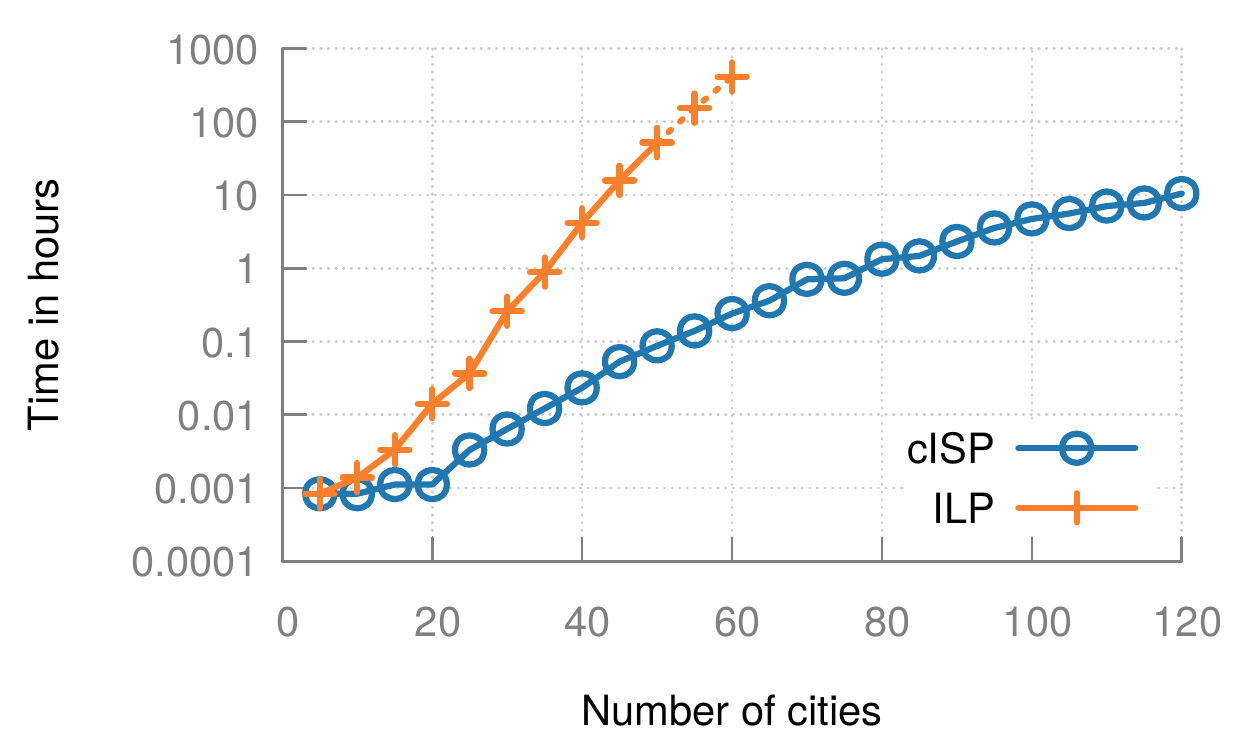}}
  \subfigure[]{\label{fig:ilp:near-optimal}
    \includegraphics[width=3in]{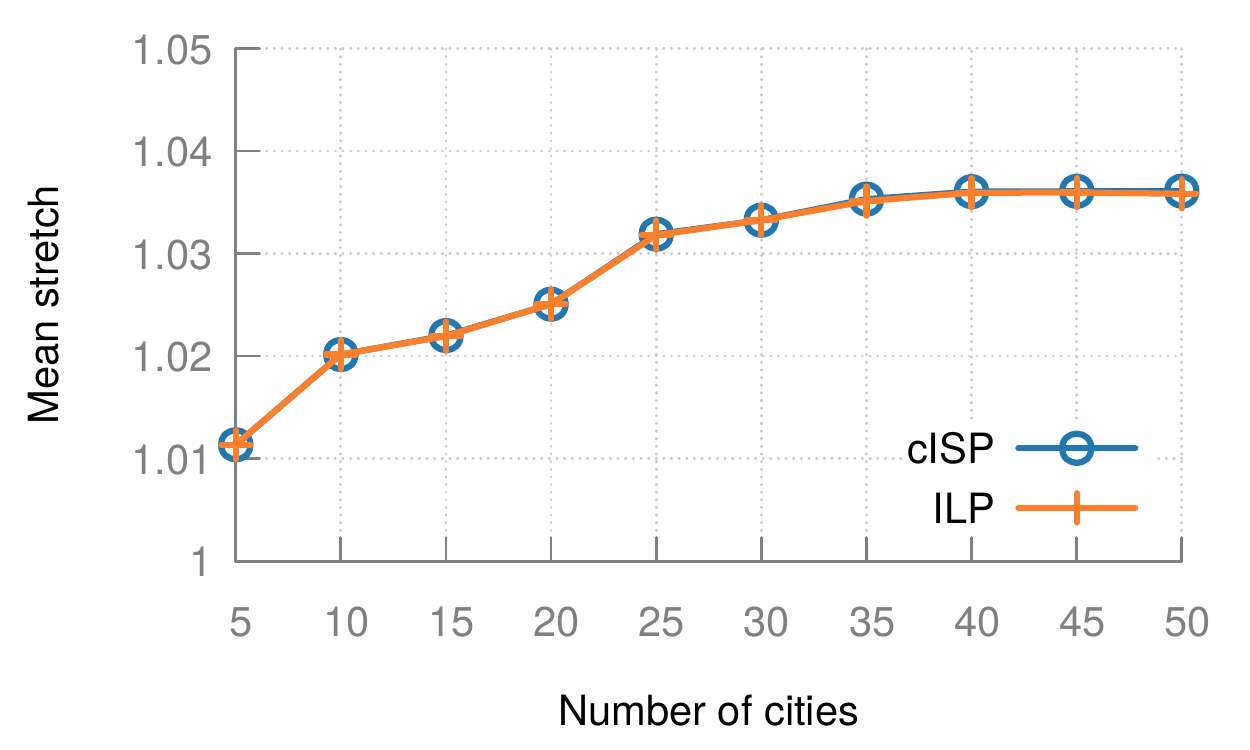}}
  \vspace{-2pt}
  \caption{\em cISP's design method is fast-enough and near-optimal: (a) cISP generates an optimized topology within hours for $120$ cities while the ILP does not yield a result even after $2$ days for more than $50$ cities. For the ILP, runtimes for $50$$+$ cities are extrapolated by curve fitting. (b) The stretch achieved by cISP matches that of the ILP to two decimal places for instances that can be optimized by the ILP.}
  \vspace{-0.1in}
  \label{fig:ilp}
\end{figure*}

\subsection{Step 3: Augmenting capacity}
\label{sec:augmenting-cap}

In many scenarios, certain links require more capacity than a single MW link provides. For short physical distances, this is a non-issue: the MW link can simply be replaced by fiber without a large impact on the network's average latency. However, for longer distances, this is not acceptable. 

One approach to resolving this problem is simply to build multiple parallel MW links, over multiple series of towers. While tower siting is often a challenging practical problem, with individual sites valued by the HFT industry at as much as \$$14$~million~\cite{oneTowerCost}, in the cISP context, there is a much larger ``tolerance'' than in HFT, where firms compete for fractions of microseconds. For a $500$~km long cISP link, the midpoint diverging $10$~km from the geodesic would increase latency by a negligible $0.2\%$. Thus, the problem of tower siting is substantially simpler. Also, in many cases, tower infrastructure is dense enough already to allow multiple parallel links. For instance, the HFT industry operates nearly $20$ parallel networks in the New York-Chicago corridor~\cite{cost}.

We can also employ a simple trick to enhance the effectiveness of parallel series of towers, as shown in Fig.~\ref{fig:bw-augment}. Instead of $k$ parallel series of towers providing merely a $k$$\times$ bandwidth improvement, connecting multiple antennae on each tower to other towers, we can obtain a $k^2$$\times$ improvement. Using antennae with overlapping frequencies requires an angular separation of $6^{\circ}$ \cite{Manning}, as shown in Fig.~\ref{fig:bw-augment}. Again, the stretch caused by the resulting gap between parallel series of towers in small. For a tower-tower hop distance of $100$~km, the minimum distance between two parallel towers should be $100 \cdot \tan(6^{\circ}) =10.6$~km, which, as noted above, has a small effect on end-to-end latency for long links.

This approach implies that for site-to-site bandwidths under $1$~Gbps, we need just one series of towers; for bandwidths between $1$-$4$~Gbps, we need $2$ series; for $4$-$9$~Gbps, $3$; etc. While tower siting circumstances are often unique, we are aided by two observations: (a) there is substantial redundancy in existing tower infrastructure, and we can often find existing towers for parallel connections (see Fig.~\ref{fig:stretchParallelPaths} and the related text in \S\ref{sec:us_cisp}); and (b) when new towers are needed, there is substantial tolerance in where they are sited, as noted above. \ankitnew{Bandwidth may potentially be increased even further through spatial diversity techniques, whereby multiple antennae are placed appropriately on the same tower
such that they can adaptively cancel interference by multiple transmission streams within the same frequency channel~\cite{winters}.}


\subsection{Generality}

Note that the above outlined approach applies broadly across other line-of-sight media, such as free-space optics and millimeter wave networking. Multiple technologies, beyond the mix of fiber and MW that we consider, can be easily incorporated into this framework, preserving relevance with technology evolution. Such line-of-sight free-space networking seems, for the near future, to be the only cost-effective solution for achieving nearly $c$-latency on the Internet.

\section{A \lowercase{c}ISP for the United States}
\label{sec:us_cisp}


We now support the above discussion of our abstract framework with a concrete instantiation: designing a cISP for the U.S. mainland. To assess line-of-sight connectivity between existing towers, we use fine-grained data on tower infrastructure, buildings, terrain, and tree canopy. The fiber conduit data is available from past work~\cite{intertubes}.

\parab{Defining the sites and traffic model:} To maximize utility while keeping costs low, we connect only the $200$ most populous cities in the contiguous United States. In addition, we coalesce suburbs and cities within $50$~km of each other, ending up with $120$ population centers. (Henceforth, when we refer to ``cities'', we refer to these population centers.) Based on population data for $2010$~\cite{populationData}, we calculate that 
$85\%$ of the US population lives within $100$~km of these $120$ cities. For the traffic matrix $H$, we set $h_{ij}$ proportional to the product of the populations of cities $v_i$ and $v_j$.

\parab{Step 1: Which city-city links are feasible?} We use existing towers listed in FCC's Antenna Structure Registration~\cite{fccasr} and databases from American Towers, Crown Castle, and several other tower companies for which we were able to download data. We cull these rather large databases of MW towers to a subset of $12$,$080$ towers as follows: \ankitnew{Towers from rental companies are typically suitable for use. From the FCC database, we only use towers over $100$~m height. When tower-density exceeds $50$ towers per $0.5$\textdegree~square grid cell, we randomly sample towers.}
(Using all towers could only improve our results, but increases compute time.)

Evaluating link feasibility across tower pairs within range of each other using the aforementioned NASA data~\cite{terrain_data}, we find $261$,$019$ tower-tower hops that satisfy line-of-sight constraints. We find that each city itself has large numbers of suitable towers in its vicinity. We run a shortest path computation on a graph comprising the cities and towers and city-tower and tower-tower hops to find the shortest city-city MW links. This yields both the cost (\ie number of towers) and latency (\ie distance along the chosen series of towers) for each city-city link.

\ankitnew{For fiber distances, we compute the shortest paths over the InterTubes~\cite{intertubes} dataset on US fiber conduits.}

\begin{figure*}
\centering
\includegraphics[width=0.95\linewidth]{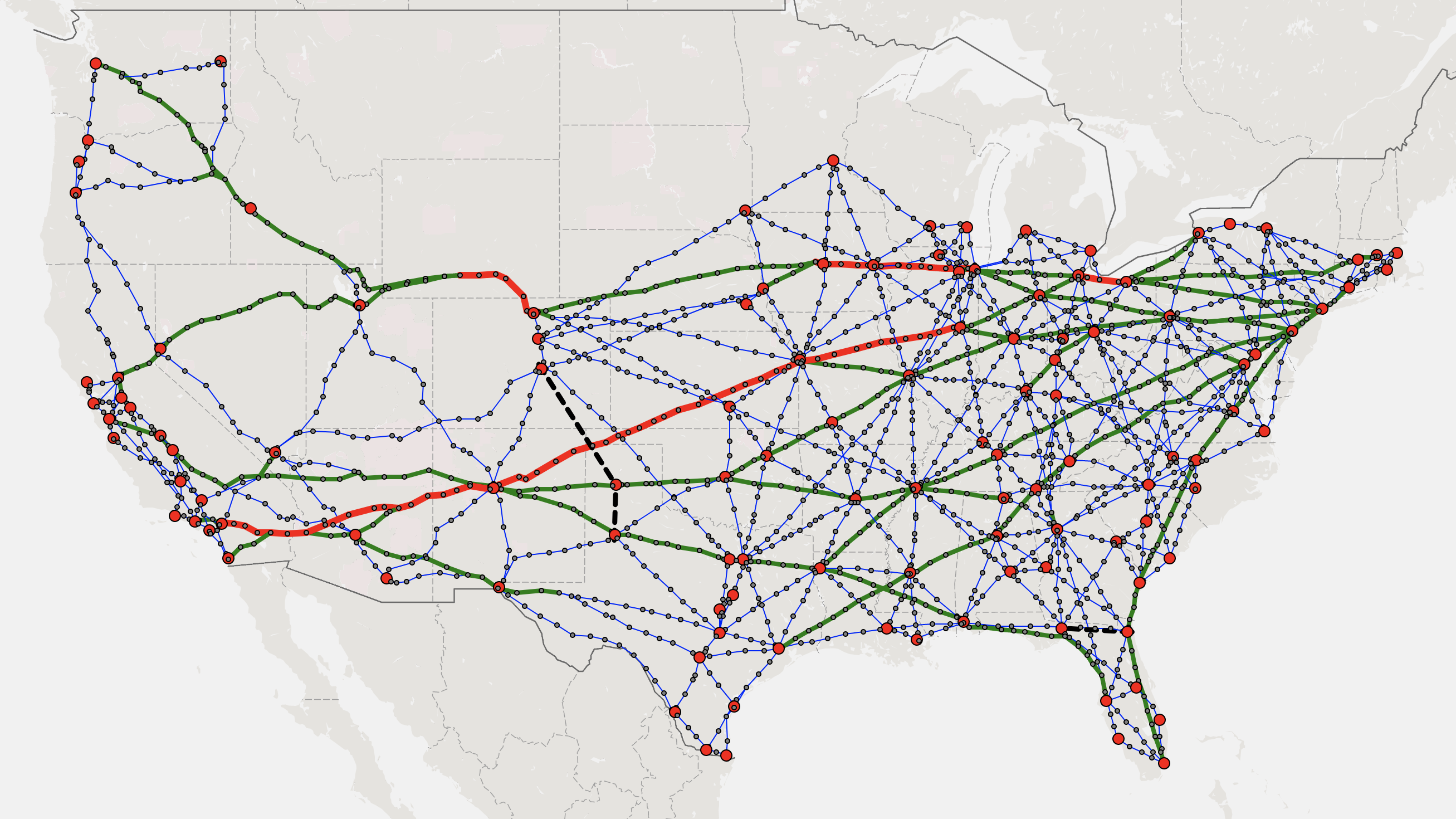}
\caption{ \em A $100$~Gbps, $1.05$$\times$ stretch network across $120$ population centers (big, red) in the US. Blue links (thin) need no additional towers. Green links (thicker) and red links (thickest) need 1 and 2 series of additional towers, respectively. The black dashed links represent fiber paths.}
\vspace{-.2in}
\label{fig:uu-100G}
\end{figure*}

\begin{figure*}[t]
  \centering
  \subfigure[]{\label{fig:towers_v_stretch}
    \includegraphics[width=2.23in]{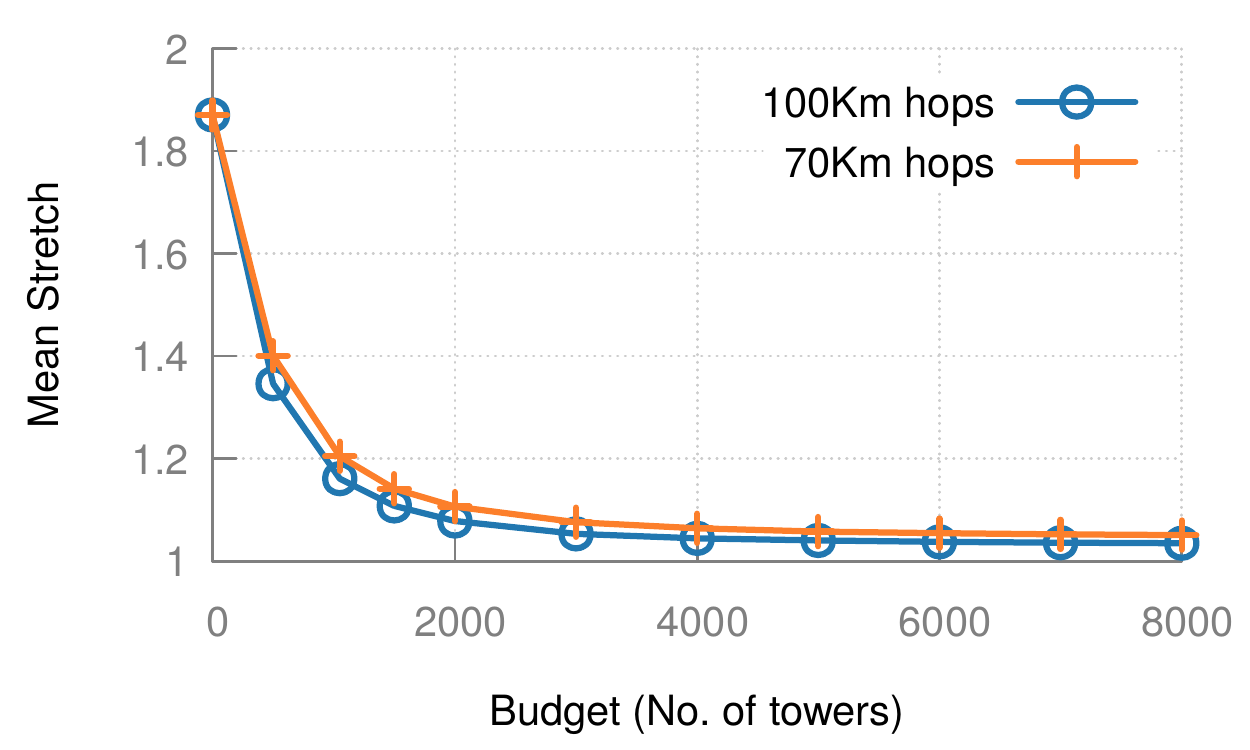}}
  \subfigure[]{\label{fig:stretchParallelPaths}
    \includegraphics[width=2.23in]{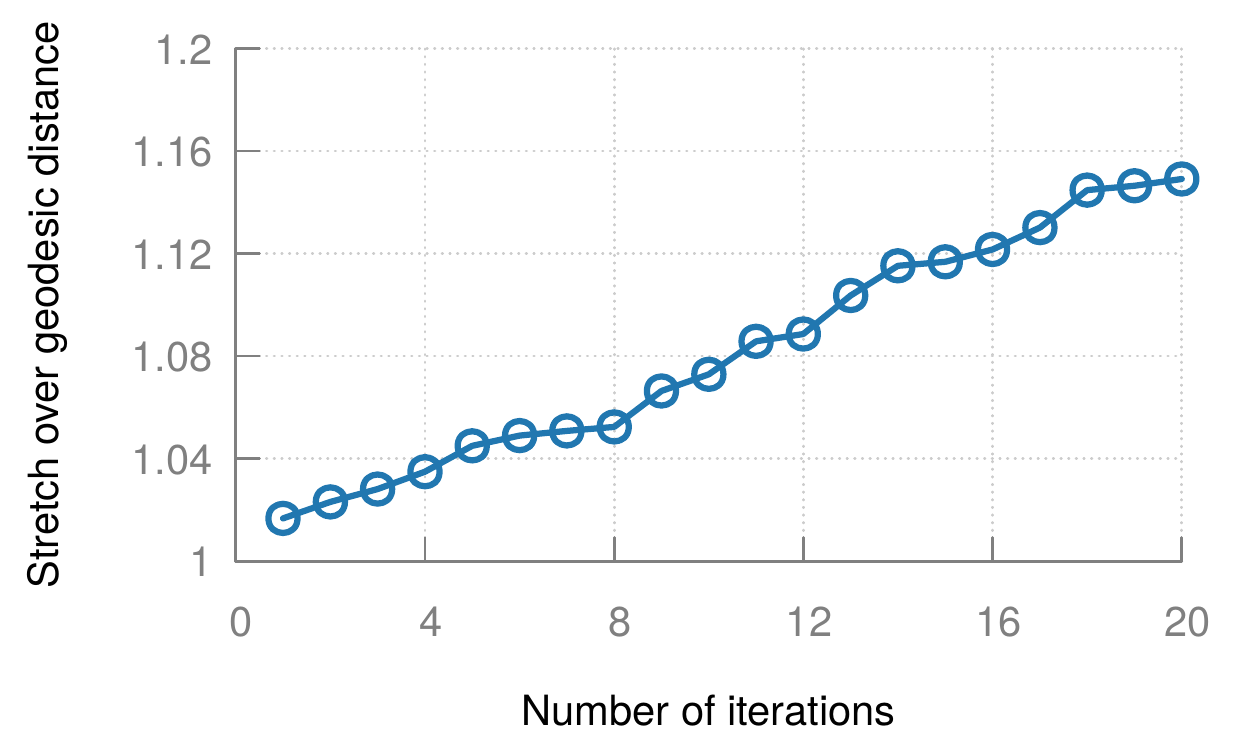}}
  \subfigure[]{\label{fig:cost_per_GB}
    \includegraphics[width=2.23in]{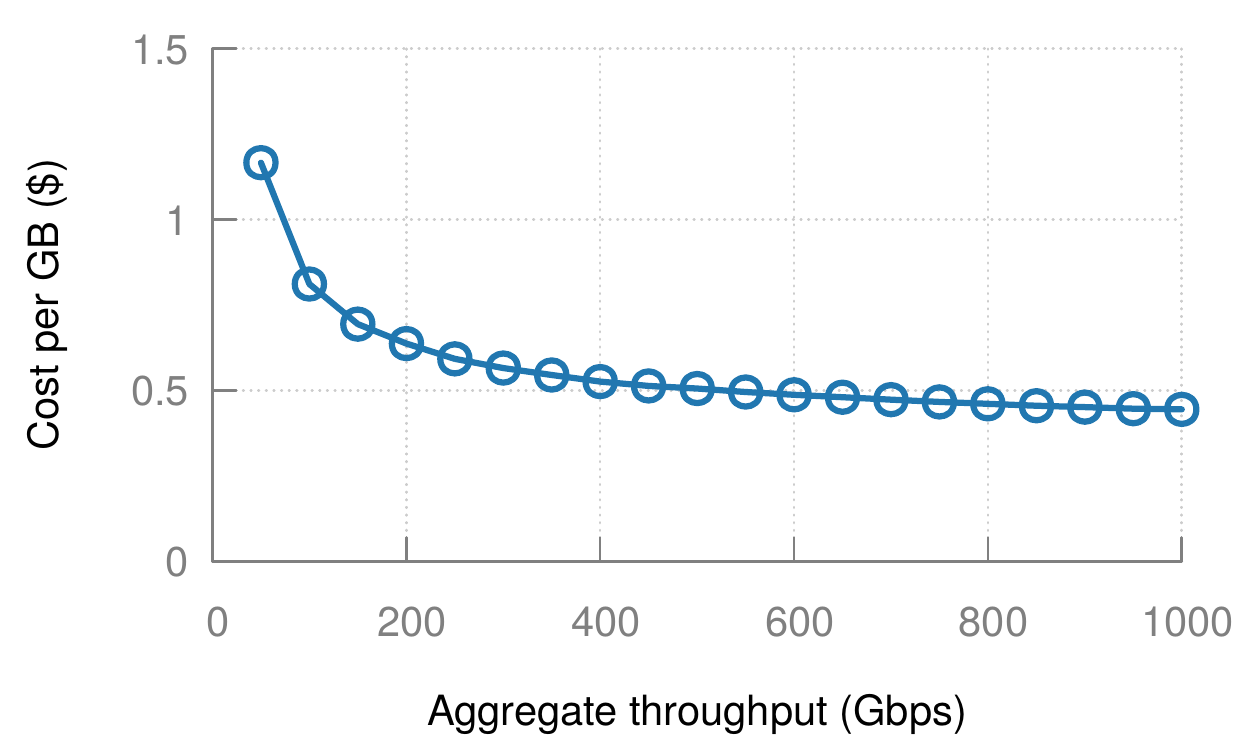}}
  \vspace{-2pt}
  \caption{\em (a) Network stretch reduces as we add more MW towers. (b) Stretch for $20$ shortest tower-disjoint purely MW paths along the long red Illinois-California link in Fig.~\ref{fig:uu-100G}. (c) Cost per GB for the city-city traffic model decreases with increasing aggregate throughput.}
  \vspace{-.2in}
  \label{fig:cisp_analysis}
\end{figure*}

\parab{Step 2: What subset of links should we build?} With the inputs now ready, we can run the algorithms of \S\ref{sec:topo-design} for any given budget to obtain a set of city-city MW links to build. We use the Gurobi solver~\cite{gurobi} for this purpose.

First, as we show in Fig.~\ref{fig:ilp:runtime}, the exact ILP, without using our observations on the problem structure, is too computationally inefficient to scale to this scenario. We use subsets of all $120$ cities to assess scalability, with the budget proportional to the number of cities in each test, with a budget of $6$,$000$ towers at the largest scale. Even after $2$ days of compute, the exact ILP was unable to obtain a result for sets of cities larger than $50$. In contrast, our cISP design heuristic is able to solve the problem at the full scale. Second, as Fig.~\ref{fig:ilp:near-optimal} shows, at small scales, where we can also run the exact ILP, our heuristic yields the optimal result. We also tested a linear program rounding approach, but even the naive LP relaxation followed by rounding did not scale beyond $60$ cities, and gave results worse than optimal. 

Fig.~\ref{fig:uu-100G} shows an example network. Designed with a budget of $3$,$000$ towers and maximum hop length of $100$~Km, its average latency is $1.05\times$ $c$-latency. Fig.~\ref{fig:towers_v_stretch} shows the reduction of the network's stretch with increases in budget for maximum hop lengths of $70$ and $100$~Kms. Given the similarities with $70$ and $100$~Kms, hereon, we only present results for the latter. An animation, showing how the network structure evolves from mostly-fiber to mostly-MW as the budget increases, is available online~\cite{anonNetworkAnim}.

\parab{Step 3: Augmenting capacity:} 
\ankitnew{We produce a target aggregate demand (\ie the sum of all site-site traffic demands) by scaling the traffic matrix $H$}. Then, each tower-tower MW hop that would be over-utilized (given the routing of \S\ref{sec:topo-design} and the $1$ Gbps capacity from \S\ref{sec:background}) is augmented with additional towers at each end, as described in \S\ref{sec:augmenting-cap}. Fig.~\ref{fig:uu-100G}'s topology, when provisioned for an aggregate throughput of $100$~Gbps, has $1$,$660$ tower-tower hops that use only already built towers seen in tower databases, while $552$ hops need one additional new tower at each end, and $86$ hops need $2$ additional towers at each end. Using the cost model described in \S\ref{sec:background}, we find that the cost per GB for this topology, with latency within $1.05$$\times$ and $100$~Gbps throughput, is \$$0.81$. For some context, this is $\sim$$10$$\times$ the cost per GB for content delivery networks~\cite{cdnCost}.

\begin{figure*}[t]
  \centering
  \subfigure{\label{fig:perturbationCompare:delay}
    \includegraphics[width=3in]{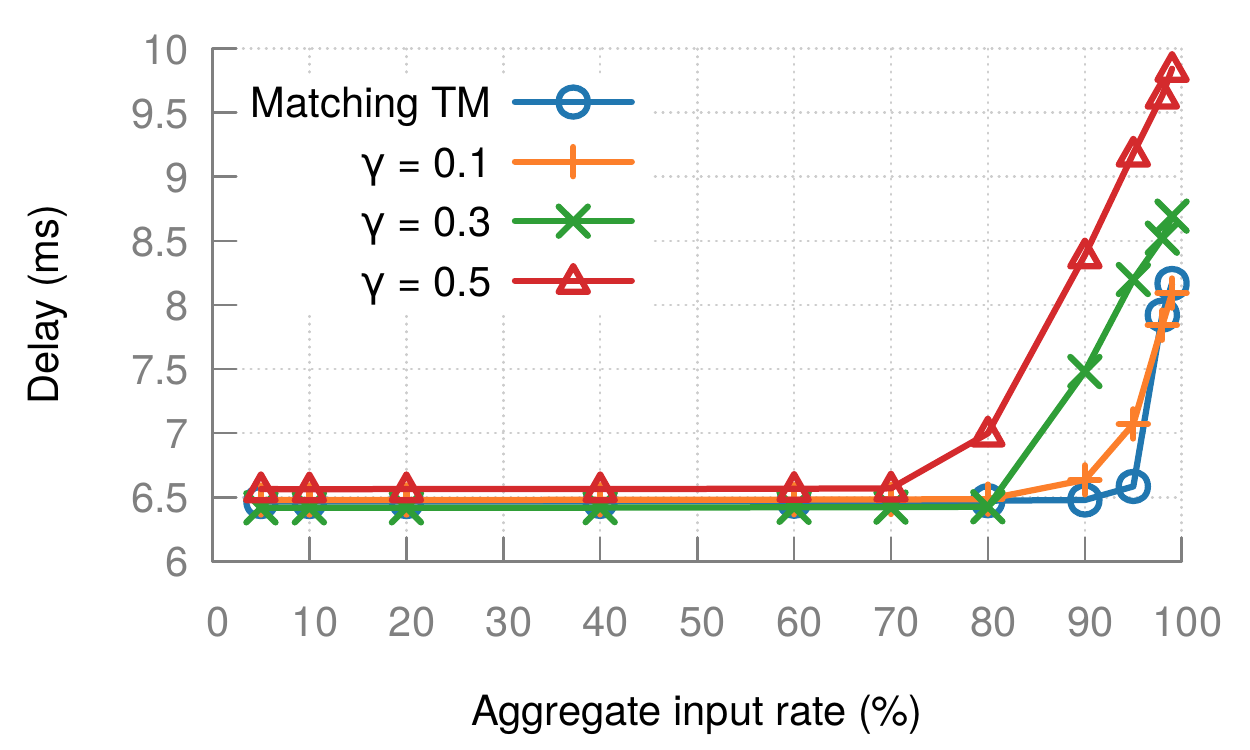}}
  \subfigure{\label{fig:perturbationCompare:loss}
    \includegraphics[width=3in]{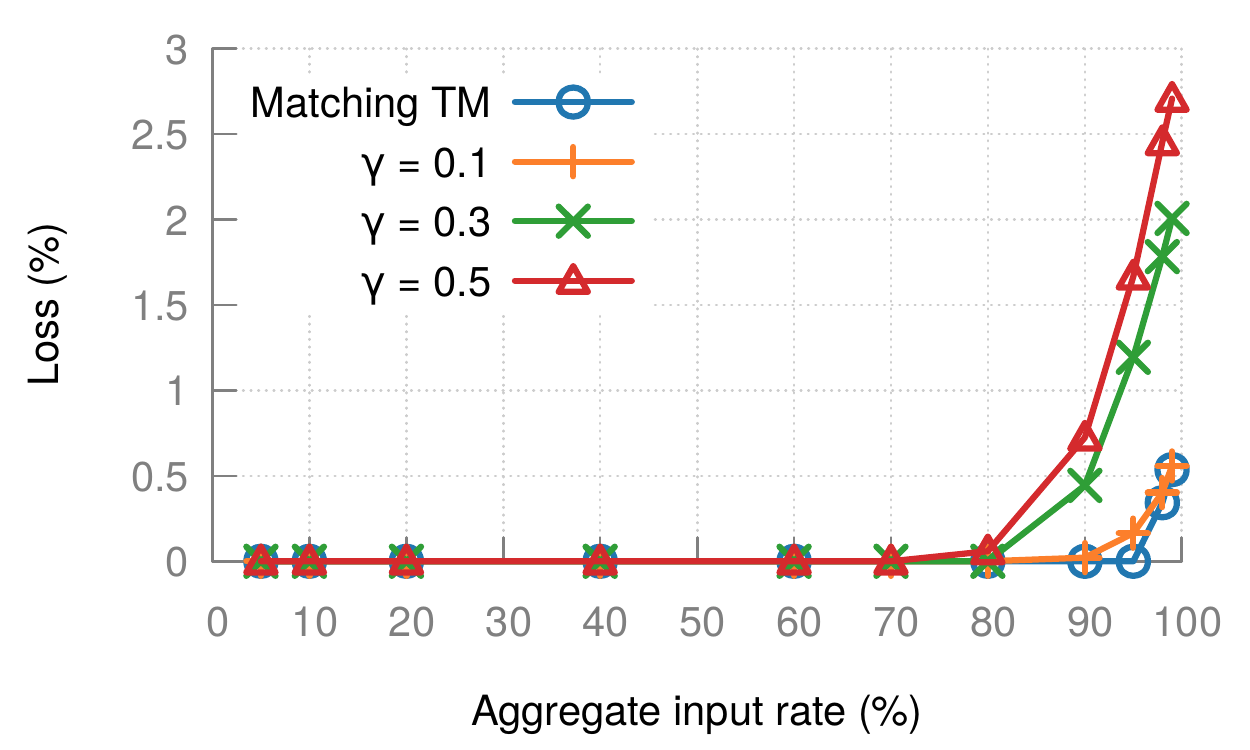}}
  \vspace{-2pt}
  \figcap{Average delay (left) and loss rate (right) remain consistent across perturbations of the city-city traffic model, except under heavy load.}
  \vspace{-.2in}
  \label{fig:perturbationCompare}
\end{figure*}

Provisioning even more bandwidth would require more new towers. 
For $1$~Tbps, some tower-tower hops
would need as many as $8$ additional towers at each end. This is not infeasible --- latency would not be inflated excessively, and towers could be found or built. In fact, for the long red link in the map in Fig.~\ref{fig:uu-100G}, which spans $2$,$700$~km from Illinois to California, we find that the \emph{longest} of these $8$ additional series of towers would be only $5\%$ longer than the shortest MW path, incurring a stretch of $1.07$, instead of $1.02$.

We can extend this argument even further: for the same Illinois to California link, we compute tower-disjoint shortest paths, \ie after finding the shortest path, we remove all towers used by it, find the next-shortest tower-path, etc. In this process, we use only existing towers from our databases, and adhere to the same link feasibility constraints. Fig.~\ref{fig:stretchParallelPaths} shows that stretch increases gradually as we keep eliminating towers; nevertheless, even after $20$ such iterations, stretch is much smaller ($1.15$) than with the existing fiber conduit ($1.75$). Note that this route runs through the Rocky mountains and other areas of low tower density. Thus, in accounting for the cost of bandwidth augmentation entirely using the (higher) cost estimates for building new towers, we are substantially overestimating the expense.

There is also another reason our costs are over-estimates: at sufficiently high bandwidth, there is a better option than building many parallel long-distance MW links: one could use the same number of towers to construct a single line of towers with shorter tower-tower distances. This can make shorter-range, but higher-bandwidth technologies like MMW or free-space optics, more cost-effective. 

Despite the above two factors, we use parallel MW towers, with all the required additional towers accounted for as new towers, to provide conservative cost-estimates as aggregate bandwidth increases in Fig.~\ref{fig:cost_per_GB}.

%% file: queuing.tex
\section{Routing \& Queuing}
\label{sec:queueing}

The HFT industry's point-to-point MW deployments demonstrate end-to-end application layer latencies within $1\%$ of $c$-latency, after accounting for all delays in microwave radios, interfacing with switching equipment and servers, and application stacks. Such low latencies across point-to-point long-distance links place sharp focus on any latencies introduced at routers for switching, queuing, and transmission.

Internet routers can forward packets in a few tens of microseconds, and specialized hardware can hit $100$$\times$ smaller latencies~\cite{routerLatency}. Transmitting $1500$~B frames at $1$~Gbps takes $12$~$\mu s$. Thus forwarding and transmission even across many long-distance links incur negligible latency. Longer routes and queuing delays, however, can have substantial impact.

To assess the impact of routing and queuing in cISP, we use ns-3~\cite{ns3}. We use UDP traffic with a uniform packet size of $500$ bytes. We use the built-in FlowMonitor~\cite{flowMonitor} to measure delay and loss rate, and add a new monitoring module to track link-level utilization. All experiments simulate $100$~Gbps of network traffic for one second of simulated time. An experiment takes approximately $10$ hours to complete on a single core of a $3.1$~GHz processor. Even achieving this running time requires some compromises: we aggregate the bandwidth of parallel links and remove the individual tower hops to focus on network links between the routing sites.





\parab{Routing schemes:} 
Besides ns-3's default shortest path routing, we implement two other schemes -- throughput optimal routing, and routing that minimizes the maximum link utilization, a scheme commonly employed by ISPs~\cite{texcp}. 

\parab{Results:} When the traffic and routing match the design target, \ie the population-product traffic routed over shortest paths, we find that the network can be driven to high utilization ($95\%$) with near-zero queuing and loss. Non-shortest-path routing schemes needlessly compromise on latency in such scenarios. (Plots for this easy scenario are omitted.)

We also test the network's behavior under deviations from the designed-for traffic model. 
We emulate scenarios where a city produces more or less traffic than expected by allowing, for each city, a ``population perturbation'' --- each city's population is re-weighted by a factor drawn from the uniform distribution $U[1 - \gamma, 1 + \gamma]$ for a chosen $\gamma \in [0,1]$.

Fig.~\ref{fig:perturbationCompare} shows the results for $\gamma\in\{0.1, 0.3, 0.5\}$. Even for large perturbations, the mean delay does not increase by more than $0.1$~ms and the loss rate is zero up to an aggregate load of $70\%$ of the capacity designed for, even with just shortest path routing. Other routing schemes are indeed more resilient to higher load, achieving virtually zero loss and queuing delay even at high utilization, but at the cost of latency. For the tested topology, both the alternative routing schemes incur $10\%$ higher latency on average (not shown in the plots).  These results indicate there would be significant value in work that reduces the amount of over-provisioning required by making modest compromises on latency on some routes, \eg as in~\cite{gvozdiev2017low}.



\begin{figure*}[t]
  \centering
  \subfigure[]{\label{fig:pacing_q}
    \includegraphics[width=3in]{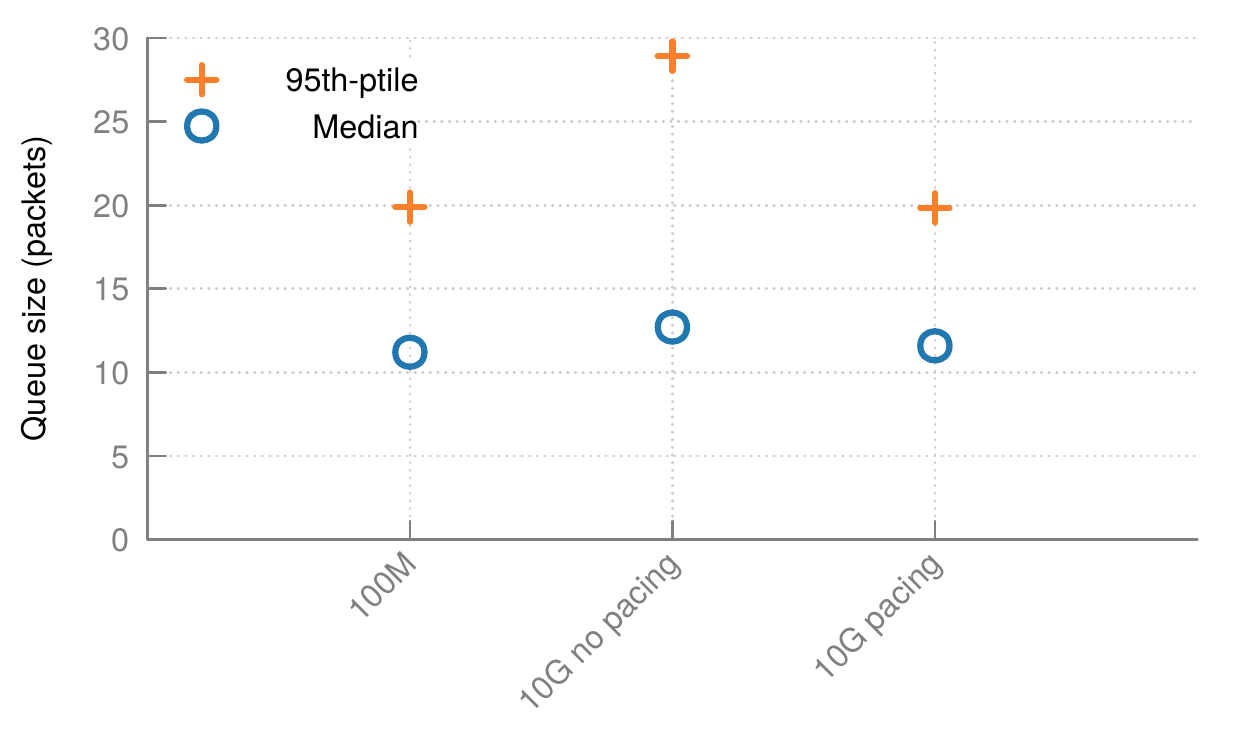}}
  \subfigure[]{\label{fig:pacing_fct}
    \includegraphics[width=3in]{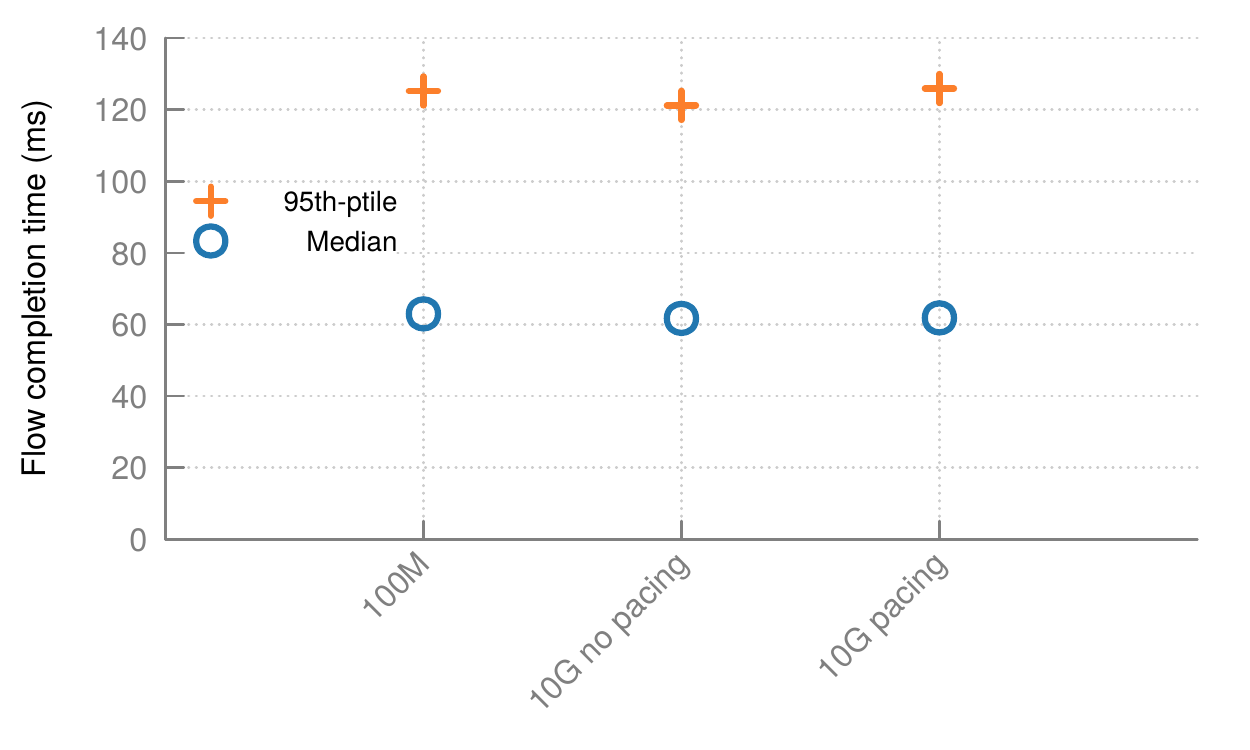}}
  \vspace{-2pt}
  \caption{\em TCP pacing addresses the problem of capacity mismatch (a) by reducing persistent queuing (b) without affecting flow completion times.}
  \vspace{-0.1in}
  \label{fig:pacing}
\end{figure*}

\parab{Speed mismatch:} The bandwidth disparity between the network core and edge for cISP may seem atypical, in the sense that in most settings, the core has higher bandwidth links compared to the edge, while in cISP, edge links (such as those at large data center end points) may often have much higher line rates when they feed their outgoing traffic into cISP. Thus, we also evaluate if this ``speed mismatch'' causes persistent congestion at cISP's ingresses.


We run ns-$3$ simulations with several sources ($S_i$) connected to a sink ($D$) through the same intermediate node ($M$). The $M$-$D$ link rate is fixed at $100$~Mbps. We then evaluate settings with every $S_i$-$M$ link being either $100$~Mbps or $10$~Gbps. The former is the control, and the latter is the setting with a speed mismatch. $M$ has an unbounded queue. Ten sources send $100$~KB TCP flows (small, as is expected in cISP) to the sink, $D$. The arrival of these TCP flows follows a Poisson process, consuming on average $70\%$ of the $I$-$D$ link's bandwidth. Each simulation run lasts $10$~s and we conduct $100$ such runs. We test TCP both with and without pacing.

Fig.~\ref{fig:pacing_q} shows that the median queue occupancy at $M$ is higher without pacing, especially at the $95$\textsuperscript{th} percentile. However with pacing, queueing behavior is nearly the same.
The median flow completion times (Fig.~\ref{fig:pacing_fct}) are unaffected both with and without pacing. 



%% file: practical.tex
\section{Practical challenges}
\label{sec:practical}

Deploying cISP would involve several practical challenges beyond network design and routing, which we now address.

\subsection{Impairments due to weather}
\label{section:weather}

\begin{figure}
\centering
\includegraphics[width=3in]{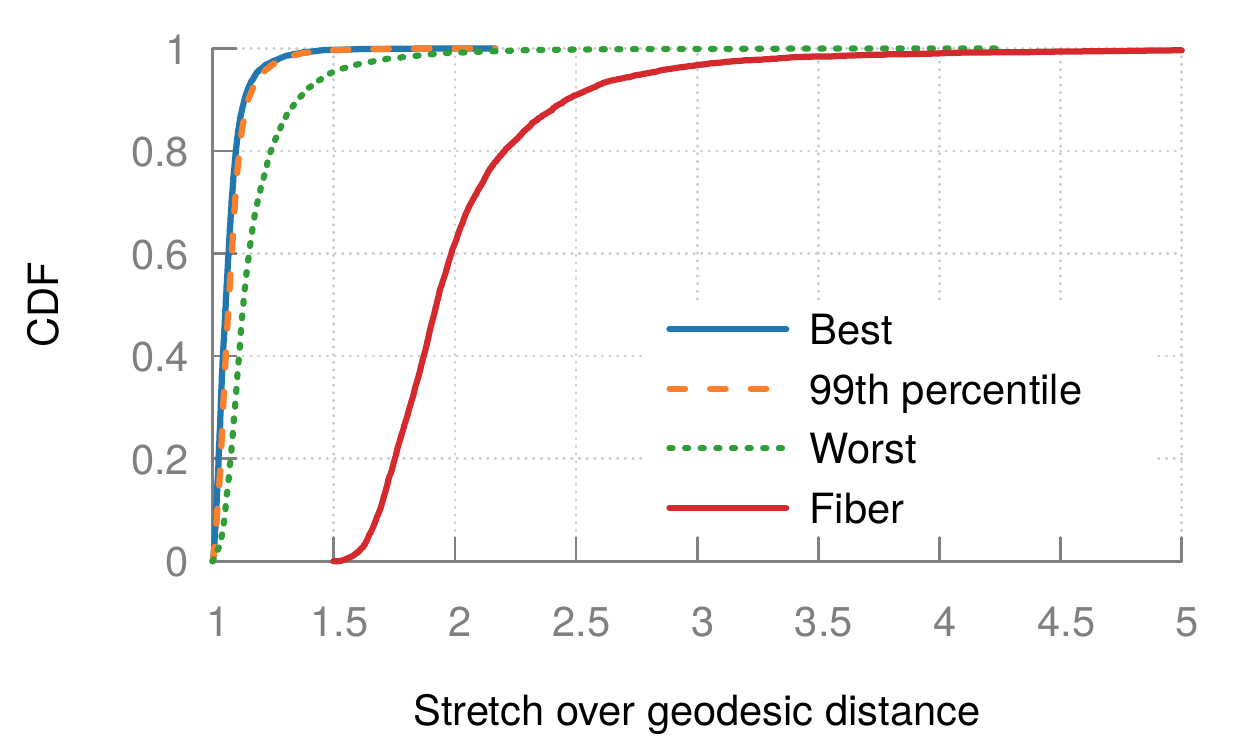}
\caption{ \em Stretch across all city-pairs over a year due to precipitation. The $99$\textsuperscript{th}-percentile stretch is comparable to the best stretch.}
\label{fig:weather_stretch}
\end{figure}

Precipitation causes MW signal attenuation. We use standard equations in MW engineering~\cite{itu_rain} to calculate attenuation. While the physical layer could trade link bandwidth for higher resilience to weather, we treat the impact of precipitation in a binary manner: if attenuation exceeds a threshold that would degrade bandwidth, we conservatively consider a link to have failed.  


We assume that when a link fails, traffic is shifted to the shortest available route, which may use any combination of MW and fiber. The high precipitation that causes failures is easy to predict, especially on the timescale of minutes. Thus, even slow, centralized management would suffice to anticipate failures and reroute accordingly.

We use NASA's precipitation data~\cite{nasa_storm_pps} to determine which links are down when, and what the impact of such failures is on the network's latency. For each day over a period of a year (July $2015$ - June $2016$), we select a $30$-minute interval uniformly at random, and identify the links that would fail during it.
We then evaluate the latency for each pair of cities end-to-end for each interval. Fig.~\ref{fig:weather_stretch} shows 
that $99$\textsuperscript{th}-percentile latencies are nearly the same as the best, fair-weather latencies. In terms of the median across city-pairs, even the worst latencies over the year are $1.7$ times lower than those over fiber. Large increases in latency due to weather typically occur only between nearby city-pairs, the fiber route to which runs through a farther-away city, \eg in Texas, Austin and Killeen fall back to a fiber route through Fort Worth.
A more sophisticated analysis allowing dynamic link bandwidth adjustment rather than binary failures can only improve these numbers. Thus, even under significantly adverse weather, most of the latency advantage of cISP remains intact.

We have also created an animated visualization of the network's latency evolving over a year's weather~\cite{anonWeatherVizLink}.


\begin{figure}
\centering
\includegraphics[width=3in]{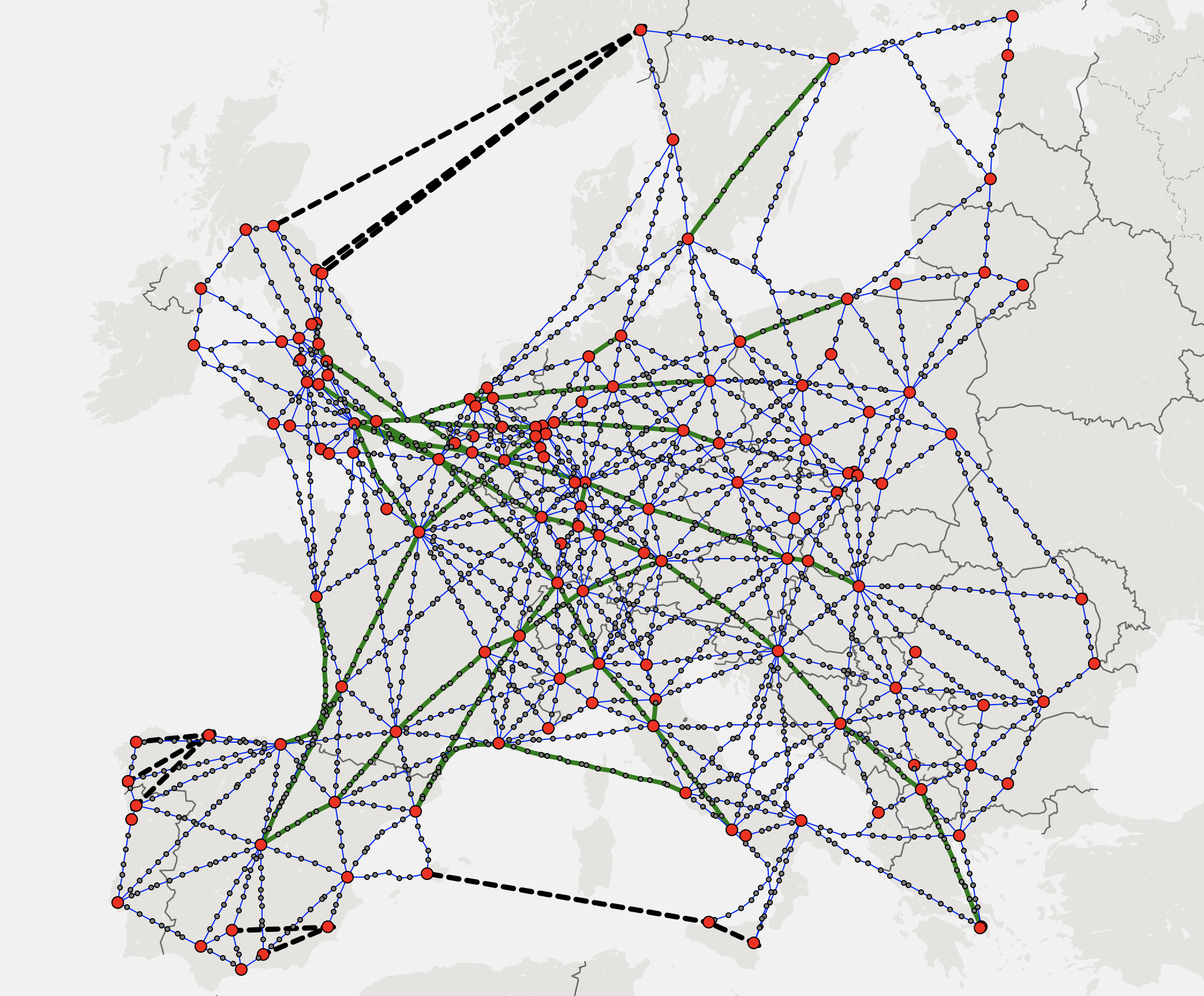}
\caption{ \em A $100$~Gbps $1.04$$\times$ stretch cISP across Europe.  This network uses several fiber connections (dashed, black lines).}
\label{fig:map_europe}
\end{figure}

\vspace*{-0.1in}
\subsection{Is the US geography special?}

So far, we have limited our analysis to the contiguous U.S. It is reasonable to ask: are the population distribution and geography of the U.S.\ especially amenable to this approach, or is it applicable more broadly? The availability of high-quality tower data and geographical information systems data for the U.S.\ enables a thorough analysis. While similar data is, unfortunately, not available to us for other geographies, we can approximately assess the design of a cISP in Europe using public, crowd-sourced data on cellular towers~\cite{openDataTower}. Lacking fiber conduit data, we assume that fiber distances between cities are inflated over geodesic distance in the same way as in the US ($\sim$$1.9$$\times$). Using our methodology in \S\ref{sec:microwaveDesign}, we design a European cISP of similar geographical scale across cities with population more than $300$k, targeting the same aggregate capacity and mean latency ($1.04$$\times$ here vs. $1.05$$\times$ for cISP-US). The cost of this design, shown in Fig.~\ref{fig:map_europe}, is similar as well, with $\sim$$3$k towers. Note that the impact of Europe's higher population density is not seen here, because we explicitly design for the same aggregate throughput. One could, alternatively, normalize throughput per capita, and compare cost per capita, to obtain similar results.

Admittedly, there is not yet a known approach to bridging large transoceanic distances using MW, limiting our approach to large contiguous land masses that need to be inter-connected with fiber. In the distant future, LEO satellite links, hollow-core fiber, or even towers on floating platforms may be of use for such connectivity.

\subsection{Is the city-city traffic model special?}

So far, our results have used the city-city population-product based traffic model. Ideally, we would be able to use wide-area traffic matrices from some ISP or content provider for modeling. In the absence of such data, we focus on showing that cISP can be tailored to vastly different deployment scenarios and their corresponding traffic models. Apart from the city-city population product model, we use (a) traffic between a provider's data centers; and (b) traffic between the cities and data centers. 

\parab{An inter data center cISP:} 
We use Google data centers
as an example, considering all $6$ publicly available US locations - Berkeley, SC; Council Bluffs, IA; Douglas County, GA;
Lenoir, NC; Mayes County, OK; and The Dalles, OR. In the absence of known inter-data center traffic characteristics, we provision equal capacity between each DC-pair.  

\parab{Data centers to the edge:} We also model a scenario where data centers are to be connected to edge locations in cities. Each of the $120$ cities connects to its closest Google data center, with traffic proportional to its population.

\begin{figure}
    \centering
   \includegraphics[width=3in]{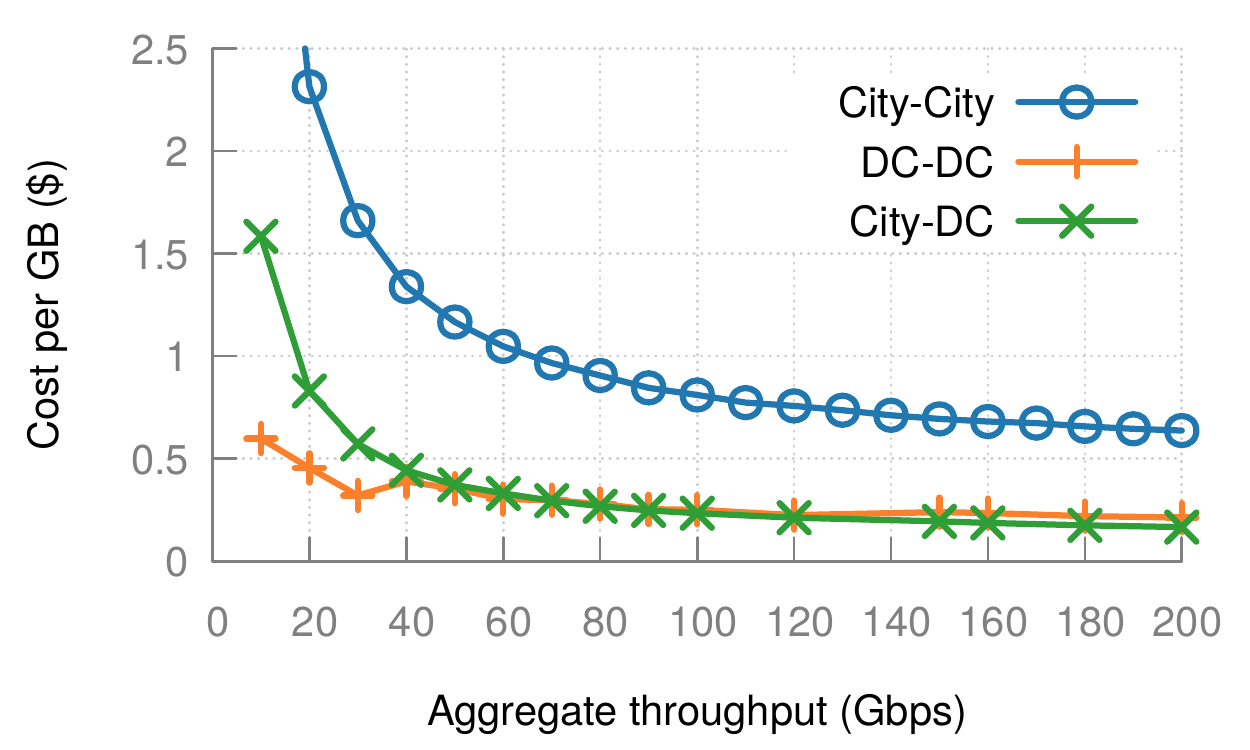}
   \caption{\em \small Cost per GB for different traffic models: the City-City model, discussed in the most detail, is the most expensive.}
    \label{fig:costPerGBtraffic}
\end{figure}

We show in Fig.~\ref{fig:costPerGBtraffic} that using the same design approach as in \S\ref{sec:microwaveDesign}, both of the above scenarios result in networks with lower cost than the city-city model. Thus, cISP can be tailored to a variety of use cases and traffic models.

\begin{figure}
\centering
\includegraphics[width=3in]{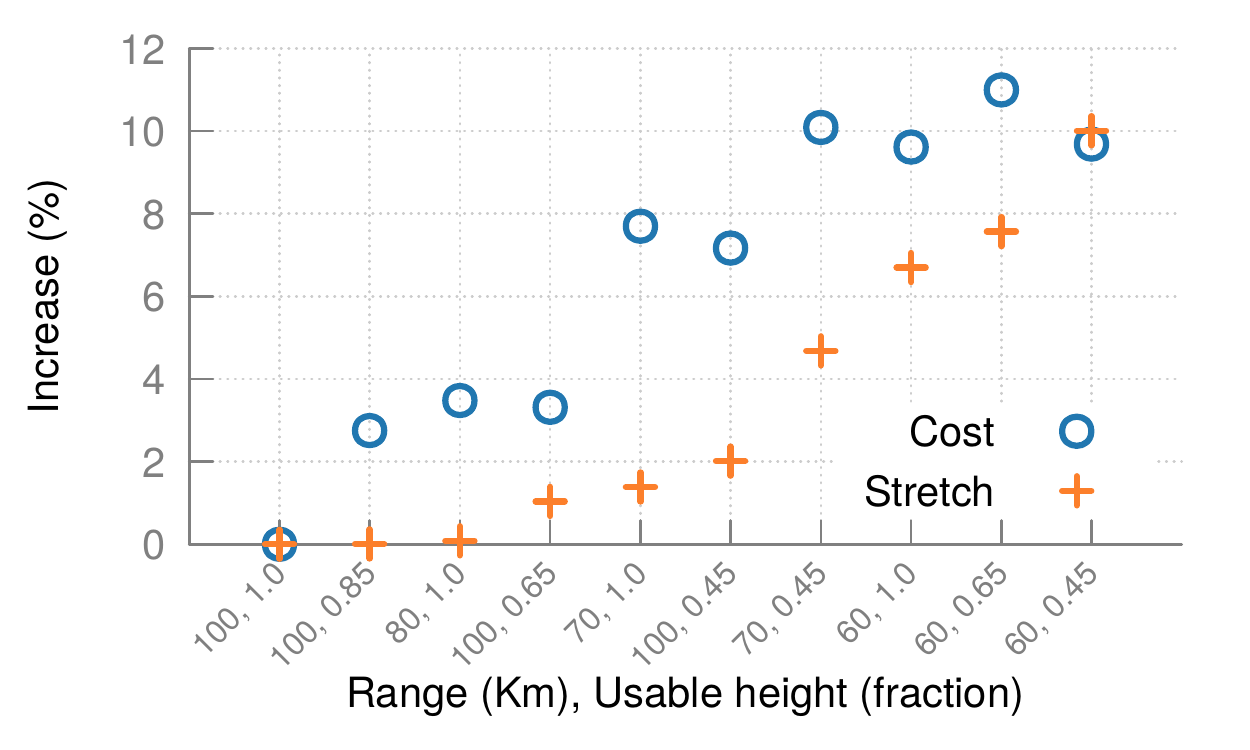}
\caption{\em \small As constraints on tower space and range become tighter, the network becomes more expensive, and stretch increases.}
\label{fig:constrained}
\end{figure}



\subsection{Traffic model mismatches}


A cISP may carry a mix of city-city, inter-DC, and DC-edge traffic. How does its performance degrade as the \emph{proportion} of these traffic types departs from the design assumptions?

We design a cISP to carry an aggregate of $100$~Gbps with a city-city : DC-edge : inter-DC traffic proportion of $4$:$3$:$3$. 
Using ns-3 simulations similar to those in \S\ref{sec:queueing}, we then test this network under several traffic mixes different from this designed-for mix --- $5$:$3$:$3$, $4$:$3$:$4$, and $4$:$4$:$3$.

\begin{figure*}[t]
  \centering
  \subfigure{\label{fig:combinedTopoCompare:delay}
    \includegraphics[width=3in]{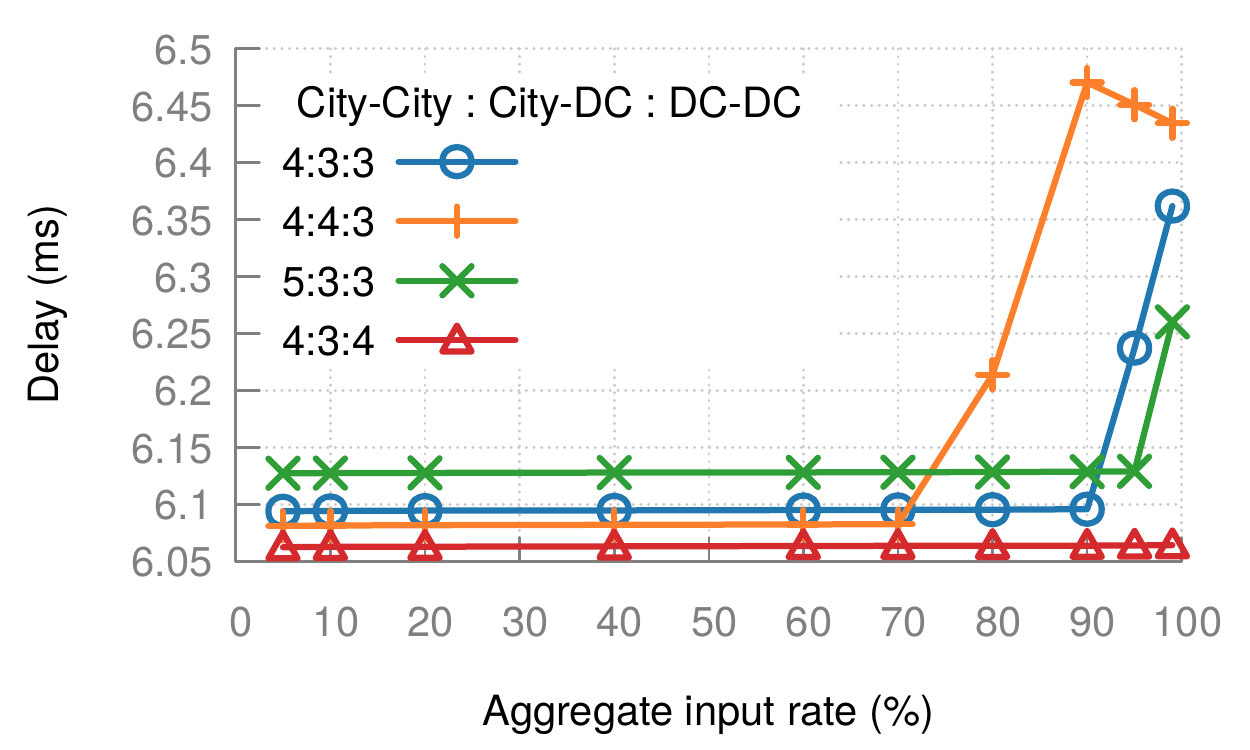}}
  \subfigure{\label{fig:combinedTopoCompare:loss}
    \includegraphics[width=3in]{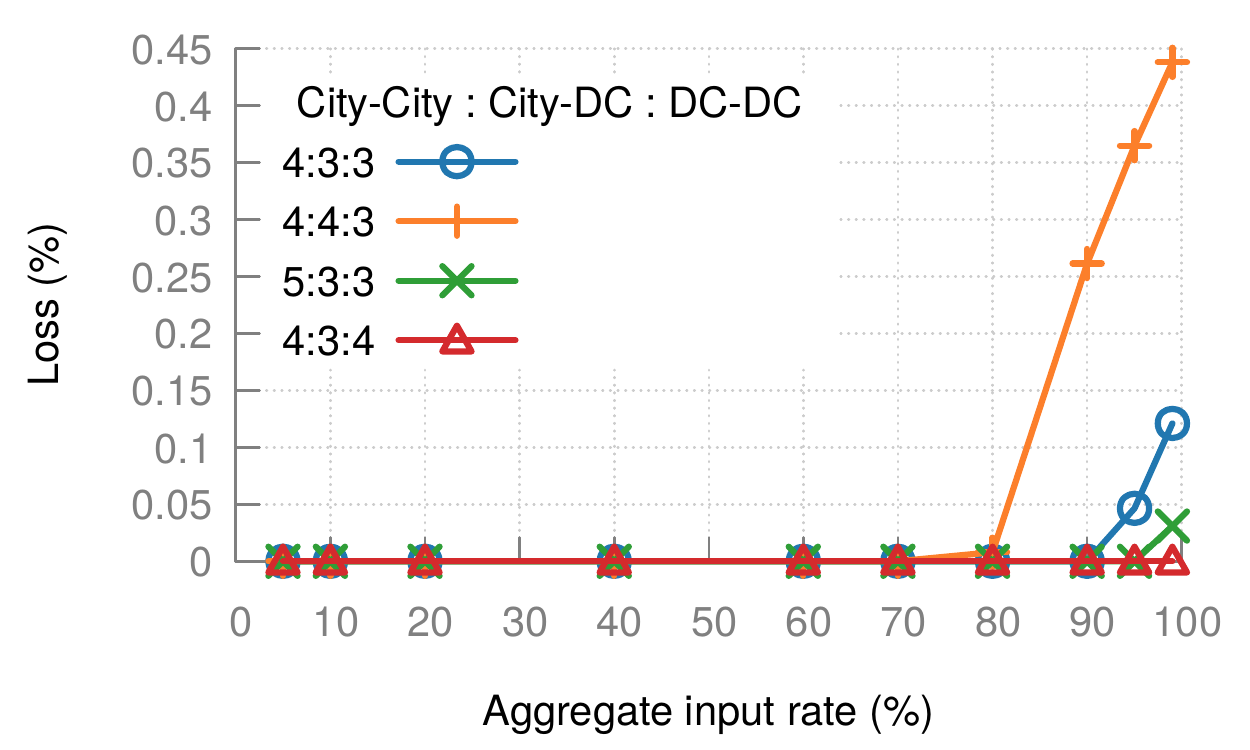}}
  \vspace{-2pt}
  \figcap{Average delay (left) and loss rate (right) remain consistent across deviations from the designed-for traffic mix, except under heavy load.}
  \label{fig:combinedTopoCompare}
\end{figure*}

Fig.~\ref{fig:combinedTopoCompare} shows that there is a difference of less than $0.05$ms in mean delay across different combinations of traffic matrices up to an aggregate load of $70\%$ of the design capacity. Similarly, loss remains nearly $0$ until this load. The decrease in delay at high load ($4$:$4$:$3$ for $x>90$ in Fig.~\ref{fig:combinedTopoCompare}) is due to losses, which are likelier on longer, higher-delay paths.

Mean delay depends more on city-city traffic, as expected: city-city traffic requires a wider infrastructure footprint, and deviations from its design parameters have greater impact. 

Thus, as discussed in \S\ref{sec:queueing}, significant traffic model deviations can be absorbed using some over-provisioning, in line with current ISP practices.

\subsection{Tower height and availability}
\label{subsec:heightAvailability}

Our initial design assumed a MW hop to be feasible if it spans a
distance of $100$~km or less, and satisfies line-of-sight constraints using the tops of the towers. In practice, however, a tower chosen for a route might not have a free spot for a new antenna at the necessary height, especially at the top, where structural concerns for large parabolic antennae are greatest, and where access and maintenance can be problematic. Further, for smaller antennae, insufficient gain margins can decrease the $100$~km maximum range. Hence, we evaluate cost and latency of the network with hop-level restrictions modeling these effects.

We test the impact of restricting usable height on towers to three levels, as a fraction of tower height: $0.85$, $0.65$, and $0.45$. Testing for line-of-sight visibility with these restrictions eliminates more towers than using tower tops. We also vary the maximum range, which can necessitate the use of a larger number of towers, thus increasing the cost and potentially making some city-pairs infeasible to connect using MW.

We assess the \emph{percentage increase} in cost and stretch values compared to the baseline values with $100$~km range and using the tower tops, \ie height fraction $=1$. Fig.~\ref{fig:constrained} shows the results for different combinations of the range and antenna-height constraints, sorted by lowest to highest stretch. The maximum increase in cost is $11\%$ (with the absolute cost per GB under these constraints being $\$0.90$), while the maximum increase in stretch is $10\%$ (with the absolute stretch compared to the geodesic being $1.16$). Thus, even substantial potential problems in tower siting and mounting antennae do not change our overall conclusions about the viability of cISP.

\gregnew{In our experience, assessments like those in this work yield accurate estimates of the latency (especially for tolerances larger than in the HFT industry) and the number of tower-tower hops that will ultimately be used to connect two sites (and hence accurate cost estimates). The precise set of towers often differs based on real-world constraints, particularly tower unavailability for structural and rental-related reasons.
Thus, while accurate in terms of cost and latency, this work does not provide fully engineered routes.} \gregnew{In practice, to improve accuracy in preparation for building a MW route, we assign each tower in a swathe connecting the sites an acquisition probability, which depends on a number of factors (\eg tower type, ownership, location). Further, for towers that can be acquired, we use a uniform distribution to model height at which space for antennae is available. With this probabilistic model, we compute thousands of candidate MW paths between site pairs, with refinements as acquisitions and height availabilities are confirmed. We make available in video form~\cite{anonRouteDemoLink} an example of such refinement.
}



\subsection{Integration into the Internet}
\label{sec:deployment}

We next discuss potential problems cISP may face in terms of integration into the present Internet ecosystem.

\parab{Low-hanging fruit:} The easiest deployment scenarios involve one entity operating a significant network backbone:

\begin{itemize}\setlength\itemsep{0pt}
    \item A CDN could use cISP for ``back-office'' traffic between its locations and content origins, which is often part of latency-sensitive user-facing interactions~\cite{pujol2014back}.
    \item Content-providers like Google and Facebook can benefit from cISP -- such WAN designs already accommodate distinctions between latency-sensitive and background traffic~\cite{swanMS,b4Google}.
    \item Purpose-built networks such as for gaming~\cite{riotGames} can easily use cISP between their edge locations and servers.
\end{itemize}

All of these are interesting and economically viable use cases with minimal deployment barriers, and each \emph{alone} may justify a design like cISP. For instance, while it is tempting to dismiss gaming as a niche, it is a large and growing market: the Steam gaming platform claims up to $16$ million players Worldwide, with $17\%$ of their traffic being US-based~\cite{steamStats}. At a $10$~Kbps rate per player,\footnote{See measurements for popular online games in~\cite{claypool2003network}.} this aggregates to $27$~Gbps -- enough to make cISP viable in this setting. (We present cost-benefit estimates, including for gaming, in \S\ref{sec:costbenefit}.)

\parab{User-facing deployment:} Access ISPs may use cISP as an additional provider, and incorporate a low-latency service into their broadband plans.\footnote{While large last-mile latencies can overshadow cISP's low latency, this is an entirely orthogonal problem, on which significant progress is being made -- 5G prototypes are already showing off sub-millisecond latencies~\cite{5Glatency}.} Utilizing cISP in this manner can help ISPs to provide and meet the requirements of demanding Service Level Agreements, the case for which was made in recent work~\cite{broadband-sla}. ISPs may use heuristics to classify latency-sensitive traffic and transit it using cISP. Alternatively, software at the user-side may make more informed decisions about which traffic may use the fast-path exposed by the ISP. While this would require significant user-side changes, note that many of today's applications already manage multi-modal WiFi and cellular connectivity. 

%% file: casestudies.tex
\section{A Few Potential Applications}

Several applications require low latency over the wide area-network. Applications focused on user interactivity, such as augmented and virtual reality, tele-presence and tele-surgery, music collaboration over long-distances, etc., can all benefit from low-latency network connectivity. Likewise, less visible and user-centric applications, such as real-time bidding for Web page advertisements and block propagation in block-chains, would also benefit from a network like cISP. While it is beyond the scope of this paper to analyze this in significant detail, we assess, in simplified environments, the improvements cISP could achieve for two application areas.

\vspace*{-8pt}

\subsection{Online gaming}
\label{sec:gaming}

We discuss cISP's benefits for both models of online gaming: \textbf{thin-client} (where each client essentially streams everything in real-time from a server) and \textbf{fat-client} (where the client has an installation of the game, performs computations, etc., and only relies on the server for updates on the game state based on other players' actions). 

Fat-clients are dominant today, and are easy to tackle: communication is almost entirely composed of latency-sensitive player actions and game-state changes, and is low-volume, typically a few Kbps per client for popular games~\cite{claypool2003network}. It can all be transferred over the low-latency network, reducing latency by $3$-$4$$\times$ compared to today's Internet.


Thin-client gaming is still in its infancy, as it depends heavily on the network, with data rates in Mbps. 
We explore the potential of a speculative execution approach: the server speculates on the game state and sends data for multiple speculated scenarios in advance over fiber, and then issues messages indicating which scenario occurred on the low-latency network. Such speculation has already shown success for rich games like ``Doom $3$'' in prior work~\cite{outatime}.


We use a toy thin-client for a multi-player Pacman variant to explore the latency benefit. 
Our rudimentary implementation speculates on all $4$ movement directions possible as user input. In line with the online-gaming literature, we measure ``frame-time'', which ``corresponds to the delay between a user's input and the observed output''~\cite{outatime}. We evaluate frame-time as latency over conventional connectivity increases (emulated by adding latency in software), and for a low-latency network always incurring $1/3$ of the latency of the corresponding conventional network.

%

\begin{figure}
\centering
\includegraphics[width=3in]{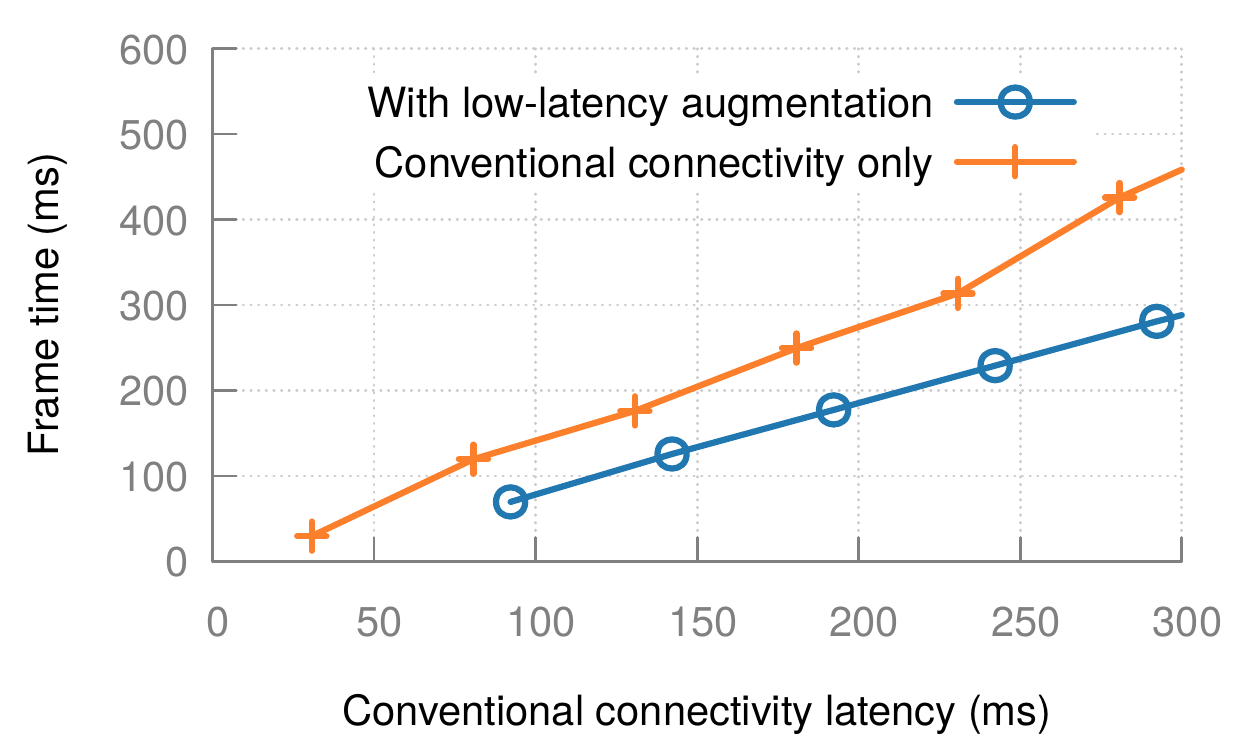}
\caption{\small \em A substantial reduction in frame time can be obtained by the use of a parallel low-latency augmentation to the present Internet.}
\label{fig:frametime}
\end{figure}

As Fig.~\ref{fig:frametime} shows, the speculative approach enabled by the low-latency network augmentation substantially reduces frame-time. This comparison would improve further if non-network overheads from processing and rendering in our naive implementation were smaller. We do not use any significant graphics on which to evaluate the additional bandwidth overhead on fiber, but even in the sophisticated scenarios examined by prior work~\cite{outatime}, this bandwidth overhead can be contained to $2$-$4.5$$\times$.

\subsection{Web Browsing}
\label{s:webeval}

\begin{figure*}[t]
  \centering
  \subfigure[]{\label{fig:page-load-times}
    \includegraphics[width=3in]{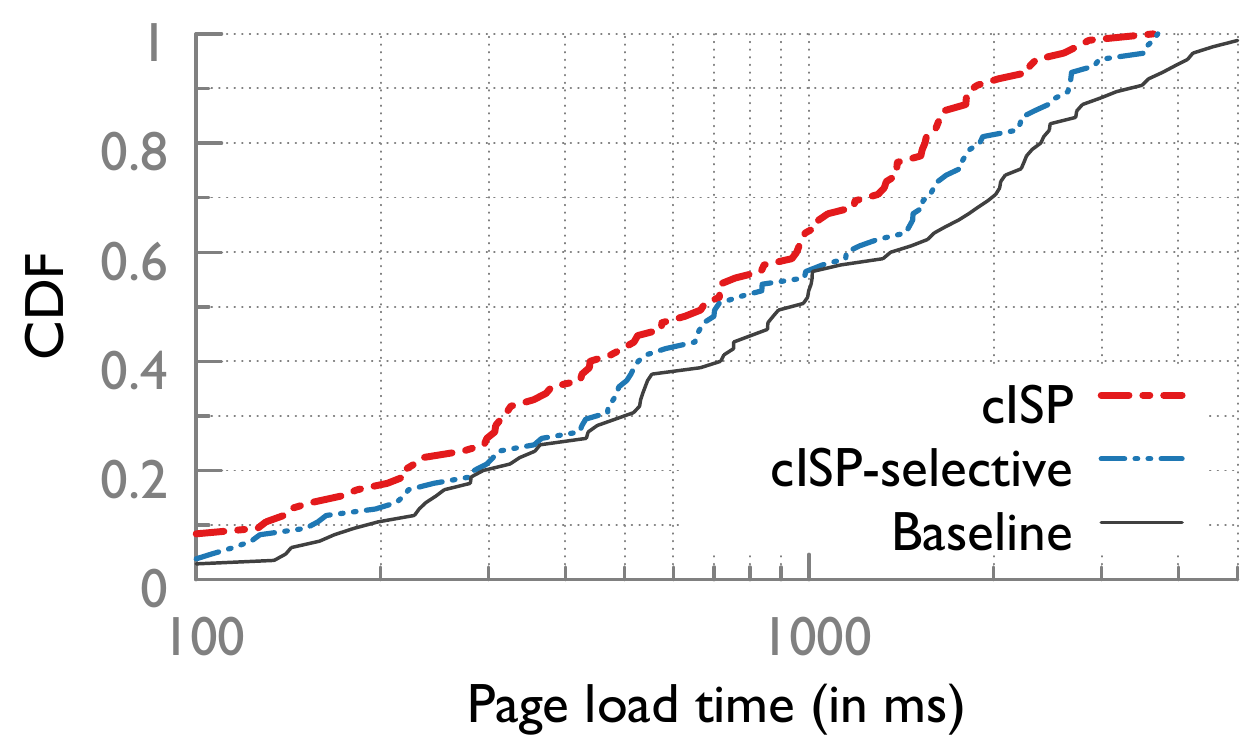}}
  \subfigure[]{\label{fig:obj-load-times}
    \includegraphics[width=3in]{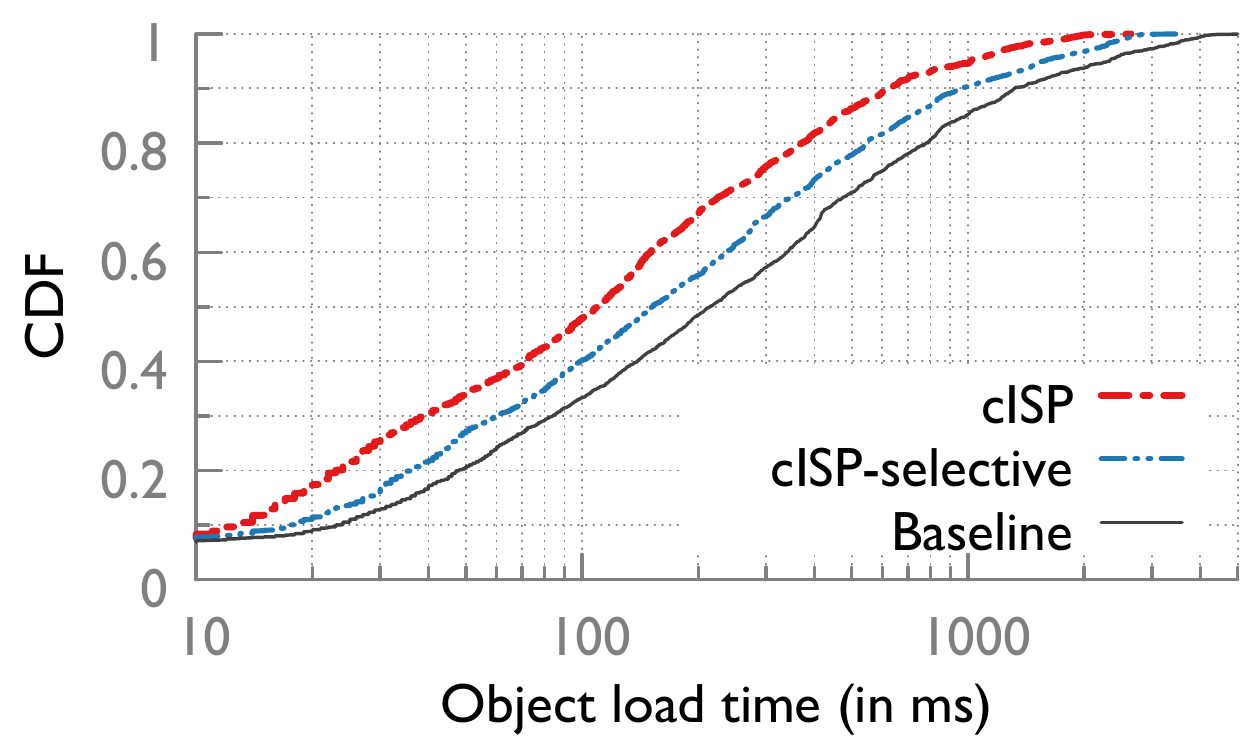}}
  \vspace{-2pt}
  \figcap{Impact of latency reductions on (a) Web page load times and (b) individual (Web page) object load times.}
  \label{fig:webeval}
\end{figure*}

We evaluate the potential impact of cISP's latency improvement on Web page load times (PLTs) (based on the {\em onLoad} event~\cite{harspec})
using Mahimahi~\cite{mahimahi}.
Our experiments use a uniform random sample of $80$ Web sites from Alexa's list of popular Web sites~\cite{alexaList}. We replay each page with unmodified latencies (as a baseline) and with latencies reduced to $0.33\times$ their original values. No bandwidth limitations are imposed.



%
%

%
%

%

Fig.~\ref{fig:page-load-times} shows the results.
Compared to the baseline, a $66\%$ reduction in latencies
(`cISP'-line) results in a $31\%$ decrease (an absolute decrease of $302$ ms) in the median PLT. This PLT reduction is
less than the $66\%$ reduction in RTT, because loading a Web page also involves significant non-network activity. For fetching the individual objects comprising these pages, these overheads are smaller, and cISP's improvements are larger. As shown in Fig.~\ref{fig:obj-load-times}, for the same $66\%$ reduction in RTT, object load times decrease by $49\%$. Small objects  (under $1460$ bytes) show a reduction of $59\%$. Thus, with a faster network like cISP, the bottlenecks shift to the protocols and Web page design.


While Web-browsing traffic comprises only a small fraction of total Internet traffic,\footnote{Cisco's 2018 estimate puts ``Web/Data traffic'' at $13\%$~\cite{ciscoForecast2017} including non-latency sensitive traffic like software updates and some file transfers.} we can further reduce the load by carrying only client-to-server traffic on cISP.  Hence, we extend Mahimahi to enable \emph{selective} manipulation of RTTs in the replay, such that some traffic sees lower RTTs than other traffic. We then emulate scenarios where only client-to-server traffic is sent over cISP at a reduced latency, \eg ``cISP-selective'' implies that only the client-to-server latencies are adjusted, and set to $0.33\times$ the recorded latency. We assume that the unadjusted latencies are symmetrical in each direction.  This approach
yields a median improvement of $27\%$ ($265$ ms) and requires sending only $8.5\%$ of the bytes over cISP.

%% file: costbenefit.tex
\section{Cost-benefit analysis}
\label{sec:costbenefit}




\ankitnew{The value of reducing Internet latencies is reflected in industry investments in this direction: Riot Games is operating its own wide-area backbone~\cite{riotGames}; Zayo acquired faster fiber routes previously used exclusively for HFT, for broader use by ``content, media and cloud providers''~\cite{zayoSpreadAcquisition}; and CDNs routinely use overlay routing to cut latency for dynamic, non-cacheable content, for which edge replication is difficult or ineffective~\cite{akamaiSureroute}.} We nevertheless present \emph{quantitative} lower-bound estimates of cISP's value per bit in a variety of contexts and assess whether its expense is justified.

\parab{Web search:} Putting together Google's quantification of the impact of latency in search~\cite{googSearch}, their estimated search revenue restricted to the US~\cite{googRevenue}, their search volume~\cite{googVolume}, estimated data transferred per search,\footnote{From Firefox desktop's network tools; mobile responses would be smaller.} and estimated cost per search~\cite{costPerSearch}, we estimate that speeding up page load times for $12$~Gbps of their US search traffic by only $200$~ms ($400$~ms) would yield an additional yearly profit of $\$87$ ($\$177$) million. This translates to an added value of $\$1.84$ ($\$3.74$) per GB.


\parab{E-commerce:} Combining Amazon.com's estimated number of visits, page fetches per visit, percentage of US traffic~\cite{amazonTraffic}, and estimated page size, we arrive at an estimate of 483 PB of traffic per year. Using their US sales estimate~\cite{amazonSales} and North America profit margin of $4\%$~\cite{amazonProfit} results in an estimated $\$7.9$ billion in profits per year. Estimates for the dependence of conversion rate on e-commerce Web sites on PLT vary from $1\%$~\cite{amazonLatency2006} to $2.4\%$ (on desktop) and $7\%$ (on mobile) per $100$~ms of additional latency~\cite{amazonLatency2017}. Thus, a speedup of $100$~ms could increase profits by $\$78.7$-$\$551$ million. If we can save $200$~ms by sending less than $10\%$ of the data over cISP (\S\ref{s:webeval}), this translates to $\$3.26$-$\$22.82$ per GB.


\parab{Gaming:} Online gamers often pay for ``accelerated VPNs'', which promise to lower network latency (perhaps using overlays). Such services cost $\$4$-$\$10$ per client per month~\cite{pingzapper,wtfast,battleping}. Full-time gaming at $8$ hours a day at a $10$~Kbps rate (as before in \S\ref{sec:deployment}) translates to $1.08$~GB / month. Thus, if cISP were priced like a cheap accelerated VPN service at $\$4$~/~mo, this would translate to a value of at least $\$3.7$~/~GB. A less aggressive model than ``full-time gaming'' would only improve cISP's value. Note that cISP's latency benefits are likely to be substantially larger than such VPN services.

Another indicator of latency's value in gaming is the market for gaming monitors with high screen-refresh rates: the $6$-$10$~ms of latency advantage is valued at over $\$50$ by many gamers, estimated from the pricing of monitors which are exactly the same except in terms of refresh rate~\cite{gamingMonitors}.

The value per GB obtained from cISP's latency reduction in above cases -- $\$1.84$-$\$3.74$, $\$3.26$-$\$22.82$, and over $\$3.7$ -- substantially exceeds its cost estimate of $\$0.81$ per GB. Even accounting for substantial over-provisioning leaves intact a clear economic argument for designs like ours. \ankitnew{Upcoming application areas like virtual and augmented reality can only make this case stronger. We expect cISP's most valuable impact to be in breaking new ground on user interactivity over the Internet, as explored in some depth in prior work~\cite{hotnetsCSpeed}.}

%% file: related.tex
\section{Related Work}
\label{sec:related}

While networking research has made significant progress in measuring latency, as well as improving it through transport, routing, and application-layer changes, the underlying infrastructure's latency inflation has received little attention, and has been assumed to be an unresolvable given. This work proposes and analyzes a nearly speed-of-light ISP, demonstrating that this is far from the case.

There are several ongoing high-profile Internet infrastructure efforts, including Google's Loon project~\cite{google-loon}, Facebook's drones~\cite{Lapowsky-WebArticle2014}, and the satellite Internet push by OneWeb and WorldVu~\cite{Allen-WebArticle2015, Geuss-WebArticle2015}. These, however, are all addressing a different problem --- expanding the Internet's coverage. One particularly noteworthy effort, from Alphabet's X moonshot factory, is a network under deployment in an Indian state, based on free-space optics, and described as ``a cost effective way to connect rural and remote areas across the state''~\cite{xFreeSpaceOptics}. Free-space networks of this type will likely become more commonplace in the future, and this work is further evidence that many of the concerns with line-of-sight networking can indeed be addressed with careful planning. Further, cISP's design approach is flexible enough to incorporate a variety of media (fiber, MW, MMW, free-space optics, etc.) as the technology landscape changes.

To the best of our knowledge, the only efforts focused on wide-area latency reduction through infrastructural improvements are in niche scenarios, such as the point-to-point links for financial markets~\cite{cost}, and isolated submarine cable projects aimed at shortening specific Internet routess~\cite{Nordrum-Spectrum2015, NEC-WebArticle2014}.

%% file: conclusion.tex
\section{Conclusion}
\label{sec:conclusion}

A speed-of-light Internet not only promises significant benefits for present-day applications, but also opens the door to new possibilities, such as \emph{eliminating the perception of wait time} in our interactions over the Internet~\cite{soslow-pam17}.

We thus present a design approach for building wide-area networks that operate nearly at $c$-latency. Our solution integrates line-of-sight wireless networking with the Internet's fiber infrastructure to achieve both low latency and high bandwidth. We use data on existing towers, and terrain and tree canopy, and a cost model reflective of current practice in engineering such networks to inform our design.

Apart from providing a near-optimal solution to the underlying network design problem, we also address numerous practical challenges, such as the availability of antenna space on towers, and assess latency degradation due to adverse weather, and deviations from the designed-for traffic model.

Lastly, our design's value far exceeds its cost for applications we could compute estimates for. Thus, greatly reducing the Internet's infrastructural latency is not only tractable, but surprisingly cost-effective, and an exciting opportunity.
